\documentclass[aps,pra,groupedaddress,showpacs]{revtex4}
\usepackage{epsfig,amssymb,amsmath}
\newtheorem{thm}{Theorem}  
\newtheorem{cor}[thm]{Corollary}
\newtheorem{lemma}[thm]{Lemma}
\newtheorem{conj}[thm]{Conjecture}

\def\wt{\widetilde}   \def\wh{\widehat}
\def\ovb{\overline}
\def\ot{\otimes} 

\def\ket{\rangle  }
\def\bra{\  \langle}
\begin{document} 
\title{Conditions for  multiplicativity of maximal 
${\ell}_p$-norms of channels for fixed integer $p$} 
\author{Vittorio Giovannetti$^1$, Seth Lloyd$^{2}$, and 
Mary Beth Ruskai$^{3}$}
\affiliation{$^1$NEST-INFM \& Scuola Normale Superiore,  
Piazza dei Cavalieri 7, I-56126, Pisa, Italy\\$^2$ 
Massachusetts Institute of Technology -- Department of Mechanical 
Engineering and Research Laboratory of Electronics \\ 77 Massachusetts Ave., 
Cambridge, MA 02139, USA
\\$^{3}$Department of Mathematics, Tufts University Medford, 
Massachusetts 02155 USA} 
\date{\today} 
\begin{abstract} 
We introduce a condition for memoryless quantum channels which, when satisfied 
guarantees the multiplicativity of the maximal ${\ell}_p$-norm with $p$ a 
fixed integer. By applying the condition to qubit channels, 
it can be shown that it is not a necessary
 condition, although some known results for qubits can be recovered. 
When applied to the  Werner-Holevo channel,
which is known to  violate multiplicativity  when $p$ is large relative to
the dimension $d$,  the condition suggests
that multiplicativity holds when  $ d \geqslant 2^{p-1}$.   This
conjecture is proved explicitly for $p=2, 3, 4$.
Finally, a new class of channels is considered which generalizes
the depolarizing channel to maps which are combinations of the 
identity channel and a   noisy one whose image is an arbitrary density matrix.   
It is shown that these channels are multiplicative for $p = 2$.
\end{abstract}
\maketitle 
\section{Introduction} 
A  noisy quantum channel can be described by \cite{Davies,KRAUS,NC} means of 
a completely-positive, trace-preserving (CPT)  map $\cal E$ 
which transforms the density matrices $\gamma$ on the Hilbert space $\cal H$
into the output states ${\cal E}(\gamma)$.  
Such maps can always be represented  \cite{Choi,KRAUS,NC}  in the form
\begin{eqnarray}
{\cal E}(\gamma)= \sum_k A_k \gamma A_k^{\dag} \;,
\qquad \sum_k A_k^{\dag} A_k = \openone \;,
\label{mappa}
\end{eqnarray} 
with $\{ A_k \}$ called a set of Kraus operators
associated with $\cal E$.   When the channel is memoryless \cite{SHOR}, $m$ successive uses of
 are described by the map ${\cal E}^{\ot m}$.
It is natural to ask if entangled inputs can decrease the effects of noise
for  memoryless channels \cite{BFS} in some way.

One measure of the effect of noise is  the maximal ${\ell}_p$-norm of a 
channel, which is defined as 
\begin{eqnarray}
\nu_p({\cal E})\equiv \sup_{\gamma  \in {\cal D}({\cal H})} 
\|{\cal E}(\gamma)\|_p \qquad p\geqslant 1\;,
\label{normach}
\end{eqnarray}
where $\|A\|_p \equiv \left( \mbox{Tr}|A|^p \right)^{1/p}$
is the $p$-norm of the operator $A$ and where the  supremum
is taken over all  ${\cal D}({\cal H})$, the set of density matrices.  
The quantity $\mbox{Tr}[{\cal E}(\gamma)^p]$   is a measure
of the  closeness of the output  to a pure state, and  $\nu_p({\cal E}) = 1$
if and only if some output state ${\cal E}(\gamma)$ is pure.    
Because the R\'{e}nyi  entropy \cite{Renyi} can be written as 
$S_p(\rho) = - \frac{1}{p-1}\log \| \rho \|_p^p$ 
one could define a maximal output R\'{e}nyi entropy \cite{AF,GIOVA1,GIOVA2}
satisfying $ (p-1) S_{p,\max}({\cal E}) = -p \log \nu_p({\cal E}) $.

Amosov, Holevo and Werner (AHW) conjectured  \cite{AHW}  that 
$\nu_p({\cal E})$ is multiplicative for tensor product channels
\begin{eqnarray}
\nu_p({\cal E}^{\otimes m})\equiv \sup_{\Gamma  \in 
{\cal D}({\cal H}^{\otimes m})} 
\|{\cal E}^{\otimes m} (\Gamma)\|_p =  \, \left[\nu_{{p}}
({\cal E})\right]^m
\label{normachmolt}\;,
\end{eqnarray}
where ${\cal E}^{\otimes m}$ is the CPT map which describes $m$ successive
memoryless uses of the channel $\cal E$, and where the maximization 
in the second 
 term of Eq. (\ref{normachmolt}) is now performed over
the density matrices
 $\Gamma \in {\cal D}({\cal H}^{\otimes m})$. 
The AHW conjecture 
requires that a product state $\Gamma$   saturates the supremum
of $\nu_p({\cal E}^{\otimes m})$ for the memoryless channel 
${\cal E}^{\otimes m}$ 
so that entangled input states $\Gamma$ do not 
increase the output norm.    One rational for the 
multiplicativity hypothesis \cite{AHW} is
the physical intuition that quantum coherence among successive channel 
uses should
be degraded by the action of a  memoryless channel. Since the ${\ell}_p$-norm 
``measures'' the purity of the states emerging from the channel, 
one might expect 
separable inputs to perform better than entangled inputs.
The multiplicativity of $\nu_p({\cal E}^{\otimes m})$ is equivalent
to additivity for the minimum R\'{e}nyi entropy with the same $p$
\cite{AF,GIOVA1,GIOVA2}.   Moreover, if (\ref{normachmolt}) holds 
for $p$ arbitrarily close to $1$, then it implies \cite{AHW} 
the additivity of the minimum 
output von Neumann   entropy \cite{KR1}, another measure of output purity.  
This has been shown \cite{SHOREQ} to be related to a conjectured additivity 
property of the Holevo information \cite{HWINFO}, and to conjectures 
about additivity and
superadditivity of the entanglement of formation \cite{AB,MSW}.

Subsequently, Werner and Holevo \cite{HW}  showed that 
the general multiplicativity
conjecture is false by producing a channel that violates
(\ref{normachmolt}) for $p>4.79$.    Nevertheless, one might still expect
multiplicativity to hold for some range of $p$, most notably 
$1 \leqslant p \leqslant 2$
and this would suffice for many applications in quantum information theory.
However, even the case $p = 2$ is still not resolved.
It is hence important to understand under which circumstances and for
which values of $p$ a 
given channel satisfies Eq.~(\ref{normachmolt}).  Many authors have  
tackled this problem by discussing special situations for which
the conjecture can be proved 
\cite{AF, AH, DHS, GIOVA1,King1, King2, King3, King4, King5, KNR, KR2,MY,SEW}.
In the  case of a fixed integer $p$,
we provide an upper bound for $\nu_p({\cal E}^{\otimes m})$, 
and derive a pair of sufficient
conditions, either of  which ensures that $\cal E$ satisfies the
multiplicativity conjecture (\ref{normachmolt}).

The material is organized as follows. In Section~\ref{s:linearization} 
we introduce some notation and present a linearization technique
that allows one to compute the ${\ell}_p$-norm of integer order
as the expectation value of an operator defined on an extended Hilbert space. 
In Section~\ref{s:suff} we derive our upper bound and show how 
it leads to a sufficient condition for the multiplicativity of the 
${\ell}_p$-norm. 
Then we apply our condition to several classes of channels.
By considering qubit channels in the case $p = 2$, we show in 
Section~\ref{s:qubits} that our sufficient condition  is not necessary.  
We also obtain new proofs of multiplicativity when the two shortest axes
of the image  ellipsoid (whether or not shifted) are equal.    
In Section~\ref{s:shft_dp} we prove multiplicativity 
when $p = 2$ for a shifted depolarizing channel and further 
generalizations which  do not seem to have been considered in the literature.
Finally, in Section~\ref{s:werner} we consider 
the Werner-Holevo channel \cite{HW} for $p = 2,3,4$, and obtain new
results about multiplicativity when $p = 3,4$.   We also conjecture that
the channel is multiplicative for {\em any} $p$ when it acts on a space
of dimension $d \geq 2^{p-1}$.

We  include several appendices.    
The first  reviews  useful facts about operators, 
including Hilbert-Schmidt duality, shift and permutation
operators, and double stochastic matrices.   Appendix A also contains  
information about the notation, and the proof of an important identity.
Appendix B discusses properties and alternative forms of the linearizing  
operators we use.   Appendix C provides details needed for our analysis
of  the Werner-Holevo channel.

\section{Linearization of $p$-norm functions}
\label{s:linearization}

\subsection{Basic linearization strategy}\label{s:single}

In this section we present a method,  introduced in \cite{GIOVA1},
that allows one to compute the ${\ell}_p$-norm 
from the expectation value of a operator defined in an extended Hilbert space.
For any integer $p$, it is possible to find a linear  
operator $X({\cal E},p)$ defined in the extended Hilbert space
${\cal H}^{\otimes p}$ such that, for 
any density matrix $\gamma \in {\cal H}$, we have
\begin{eqnarray}
\mbox{Tr} \big[{\cal E}(\gamma) \big]^p = 
\mbox{Tr}[\,( \underbrace{\gamma\otimes\gamma \otimes 
\cdots \otimes \gamma}_{p \mbox{\small{-times}}}) \; X({\cal E},p)\,] 
\label{prop1}
\end{eqnarray}
where the trace in the left-hand side is computed with respect to 
an orthonormal basis of $\cal H$, while the trace in the right-hand side
is computed with respect to an orthonormal basis of ${\cal H}^{\otimes p}$.
In other words, we can represent the $p$-purity function 
$\mbox{Tr}[{\cal E}(\gamma)]^p$ as the
expectation value of  
$X({\cal E},p)$ on $p$ copies of $\gamma$.  
The operator $X({\cal E},p)$ is
not uniquely defined; in fact, it can be realized by the
action of tensor products of the dual map of $ {\cal E}$ 
on any permutation operator acting on ${\cal H}^{\otimes p}$
whose shortest cycle is length $p$.

To make this explicit, we need some notation, which is
explained in more detail in Appendix~\ref{app}, particularly
sections \ref{s:dual} and \ref{s:shift}.
We will use a hat to denote the dual, or adjoint, map
$\wh{{\cal E}}$ with respect to the Hilbert-Schmidt
inner product.  
Let   $L_p $ and $R_p$ denote the left and right cyclic shifts
which can be defined by their action on an orthonormal
product basis as
 \begin{subequations}  \label{shft} \begin{eqnarray}    
  L_p  | \xi_1 \xi_2 \cdots \xi_{p-1} \xi_p \rangle & = & 
   | \xi_2 \cdots  \xi_p  \xi_1 \rangle \\
   R_p | \xi_1 \xi_2 \cdots \xi_{p-1} \xi_p \rangle & = & 
   | \xi_p \xi_1 \cdots  \xi_{p-1}   \rangle.
\end{eqnarray} \end {subequations} 
where $ | \xi_1 \xi_2 \cdots \xi_{p-1} \xi_p \rangle = 
  \ot_{j=1}^p  | \xi_j   \rangle $ and
   $\{  | \xi_k \rangle \}$ is an orthonormal basis for ${\cal H}$.  
Then the operator 
\begin{eqnarray}  \label{omega}
\Omega({\cal E},p) \equiv  \wh{\cal E}^{\otimes p}(L_p )
\end{eqnarray}
satisfies (\ref{prop1}).    
This follows from
\begin{eqnarray} \label{omegaprop}
\mbox{Tr} \,  \gamma^{\otimes p}  \, \Omega({\cal E},p)
  & = &   \mbox{Tr} ( \gamma\otimes\gamma \otimes  \cdots \otimes \gamma )  
\,  \wh{\cal E}^{\otimes p}(L_p )  \nonumber \\  & = &
\mbox{Tr} \,    \big[ {\cal E}(\gamma) \otimes {\cal E}(\gamma) \otimes 
\cdots \otimes  {\cal E}(\gamma) \big]  L_p   \nonumber \\
& = & \mbox{Tr}[{\cal E}(\gamma)]^p    
\end{eqnarray}
where the last step used  (\ref{lemma}).   It follows  from (\ref{lemma1}) 
that  $L_p$ could be replaced
by another permutation; however, it is important to make a
definite choice for later use.

In previous work  \cite{GIOVA1,GIOVA2}, a different realization of 
$X({\cal E},p)$ was used which is valid only for pure states.   
Let
\begin{eqnarray}
\Theta({\cal E}, p) & = &  \Omega({\cal E},p) \, R_p = 
\wh{\cal E}^{\otimes p} (L_p ) \; R_p 
\label{theta}  \\
& = & \sum_{k_1,\cdots,k_{p}}  
A^{\dag}_{k_1} A_{k_2} \otimes A^{\dag}_{k_2} A_{k_3}
\otimes \cdots \otimes A^{\dag}_{k_p} A_{k_1} 
\label{thetafin} 
\end{eqnarray}
where $\{ A_k \}$ form a set of Kraus operators for ${\cal E}$ 
as in (\ref{mappa}). The operator $ \Theta({\cal E}, p)$ 
satisfies~(\ref{prop1}) when $\gamma = |\psi \rangle \langle \psi |$ 
is a pure state.
This relation is proved in Appendix~\ref{s:thetapf}, and 
implicitly shows that it does not depend on the chosen 
Kraus representation~(\ref{mappa}) of $\cal E$.
For $p=2$, (\ref{thetafin}) and (\ref{theta}) were  
obtained earlier in Ref.~\cite{ZANNA}. 

In general, the operator $\Omega({\cal E},p)$ will not be Hermitian.
We have already observed that 
$X({\cal E},p)$ is not unique and that whenever $P_p$ 
is a permutation operator 
whose shortest cycle is length $p$, the operator
$\wh{\cal E}^{\otimes p}(P_p)$ provides another realization.  
Since $L_p = R_p^{\dagger}$, 
\begin{eqnarray}
\big[ \wh{\cal E}^{\otimes p}(L_p ) \big]^{\dagger} =  
\wh{\cal E}^{\otimes p}(L_p ^{\dagger})
= \wh{\cal E}^{\otimes p}(R_p).
\end{eqnarray}
This implies that $\Omega({\cal E},2)$ is Hermitian for $p = 2$, and 
that the operator  
\begin{eqnarray}  \label{Herm}
\tfrac{1}{2}
\big[ \Omega({\cal E}, p) +  [ \Omega({\cal E}, p)]^{\dagger} \big] =
\tfrac{1}{2} \big[ \wh{\cal E}^{\otimes p}(L_p ) +  
\wh{\cal E}^{\otimes p}(R_p) \big]
\end{eqnarray}
gives a Hermitian realization of $X({\cal E},p)$ for any $p$. 
However, we do not expect (\ref{Herm}) to  have the important
multiplicity property (\ref{ellemme}) for repeated uses
of the channel.
Further discussion of other realizations $X({\cal E},p)$ is given
Appendices~\ref{s:Xperm} and \ref{s:Xpure}.

Linear operators satisfying (\ref{prop1}) provide a useful tool 
for studying the  $p$-purity functions, 
 which are intrinsically non-linear objects;
it reduces some associated problems to the  
analysis of the linear operator $X({\cal E},p)$
acting on the extended Hilbert space ${\cal H}^{\otimes p}$ 
obtained by adding $p-1$ ``fictitious''
copies of the input Hilbert space $\cal H$. 
In Refs. \cite{GIOVA2,GIOVA1}, this approach
was used to obtain some additivity
properties of Gaussian Bosonic channels. 
For $p=2$, Eq.~(\ref{prop1}) was used
in Ref. \cite{CAVES} to study the fidelity obtainable in continuous-variable 
teleportation with finite two-mode squeezing, and in Ref. \cite{ZANNA}
to analyze the purity of generic quantum channels. 

\subsection{Tensor product maps}\label{s:multi}
The results derived in the preceding section can also be
applied when the basic CPT map is itself a  tensor
product.    Then Eq. (\ref{prop1}) becomes
\begin{eqnarray}
\mbox{Tr} \big[{\cal E}^{\otimes m} (\Gamma)\big]^p  = 
\mbox{Tr}[\,( \underbrace{ \Gamma\otimes \Gamma \otimes 
\cdots \Gamma }_{p\mbox{\small{-times}}}) \; X({\cal E}^{\otimes m},p)\,] 
\label{prop2}\;,
\end{eqnarray}
where  $\Gamma$ is a generic 
density matrix in the input Hilbert space ${\cal H}^{\otimes m}$
and $X({\cal E}^{\otimes m},p)$ is a linear operator
on $\big({\cal H}^{\otimes m} \big)^{\otimes p} = {\cal H}^{\otimes mp}$. 
Following the strategy of Section~\ref{s:single},  we now choose
$X({\cal E}^{\otimes m},p)$ to be the operator,
\begin{eqnarray}
\Omega({\cal E}^{\otimes m},p) \equiv  
\big( \wh{{\cal E}^{\otimes m}}\big)^{\otimes p} ({\mathbb L}_p) 
= \big( \wh{{\cal E}^{\otimes m}}\big)^{\otimes p} (L_p^{\otimes m}) 
\end{eqnarray}
The operator ${\mathbb L}_p$ is described in more detail in 
Appendix~\ref{s:shiftmp} where it is proved that
${\mathbb L}_p = L_p^{\otimes m} = (L_{mp})^m$.  Using 
$\big(\wh{{\cal E}^{\otimes m}}\big)^{\otimes p} = 
\big(\wh{{\cal E}}\big)^{\otimes mp}$, we find
\begin{eqnarray}
\Omega({\cal E}^{\otimes m},p) =
 \wh{{\cal E}}^{\otimes m p} (L_p^{\otimes m}) 
= \left[ \wh{{\cal E}}^{\otimes p} (L_p) \right]^{\otimes m}
= \big[ \Omega({\cal E},p) \big]^{\otimes m} .
\label{ellemme}
\end{eqnarray}

Equation (\ref{ellemme}) is a key result whose simplicity hides a great
deal of subtlety.    The essential point is that the linear operator  
$X({\cal E}^{\otimes m},p)$ which satisfies (\ref{mappa})  for
the tensor product channel  ${\cal E}^{\otimes m}$  can be realized by the 
action of the dual
of ${\cal E}^{\otimes m}$ on the permutation $L_p^{\otimes m}$.

\bigskip
% \pagebreak

\section{Conditions for  multiplicativity}\label{s:suff}

\subsection{Upper bound}\label{s:upper}
We now use the singular value decomposition \cite{HJ1,SVD}
to observe that one can write
\begin{eqnarray}  \label{eqSVD}
\Omega({\cal E},p) = \sum_j \mu_j \, | \eta_j \rangle \langle \omega_j|
\end{eqnarray}
where $\{ | \eta_j \rangle \}$ and $\{ | \omega_j \rangle \}$ 
denote orthonormal bases for  ${\cal H}^{\ot p}$
and $\mu_j > 0 $ are the singular values of $\Omega({\cal E},p)$,
i.e., the non-zero eigenvalues of
 $|\Omega({\cal E},p)| \equiv  \sqrt{[\Omega({\cal E},p) ]^{\dagger} 
\Omega({\cal E},p)}$.  
 Before applying this, it is convenient to introduce the convention
of using bold uppercase Greek letters to denote tensor product vectors as in
   $|\boldsymbol 
\Psi \rangle \equiv  |\psi\rangle^{\ot p} =  \linebreak
|\psi\rangle \ot |\psi\rangle \ot \cdots \ot |\psi\rangle 
\in {\cal H}^{\ot p}$.  Then
 \begin{eqnarray}
   \mbox{Tr} \big[{\cal E} \big(|\psi\rangle\langle \psi| \big) \big]^p  & = &
\langle \boldsymbol{\Psi} | \Omega({\cal E},p)
|\boldsymbol \Psi \rangle  \nonumber \\
&= &\sum_j  \mu_j
\langle \boldsymbol{\Psi} | \eta_j \rangle 
   \langle \omega_j  | \boldsymbol{\Psi} \rangle  \nonumber \\
   & \leq & \mu_{\max} \sum_j 
\big| \langle \boldsymbol{\Psi} | \eta_j \rangle 
   \langle \omega_j  | \boldsymbol{\Psi} \rangle \big| \\
& \leqslant &  \mu_{\max} \|  \boldsymbol{\Psi} \|^2
   =  \| \Omega({\cal E},p) \|_{\infty}  \nonumber
 \end{eqnarray}
where $\mu_{\max}= \sup_j \mu_j   = \| \Omega({\cal E},p) \|_{\infty}  $
is the largest singular value of  $\Omega({\cal E},p) $. 
Applying this analysis to multiple uses of the channel, one
can similarly  conclude that
\begin{eqnarray}  \label{bndm}
\mbox{Tr}\big[{\cal E}^{\otimes m} \big(|\Psi\rangle\langle \Psi|
\big) \big]^p & \leqslant  & \|  \Omega({\cal E}^{\otimes m},p) \|_{\infty},
\end{eqnarray}
where $| \Psi \rangle$ is now an arbitrary vector in ${\cal H}^{\ot m}$.
However, it follows from (\ref{ellemme}) that the singular values
of $\Omega({\cal E}^{\otimes m},p)$ are products of those of 
$\Omega({\cal E},p)$
so that 
\begin{eqnarray}   \label{svalm}
 \|  \Omega({\cal E}^{\otimes m},p) \|_{\infty} = 
    (\big \| \Omega({\cal E},p) \|_{\infty}  \big)^m = (\mu_{\max})^m.
\end{eqnarray}
Combining (\ref{bndm}) and (\ref{svalm}), one finds
\begin{eqnarray}  \label{bndfin}
\mbox{Tr}\big[{\cal E}^{\otimes m} \big(|\Psi\rangle\langle \Psi|
\big) \big]^p  \leqslant   
    (\big \| \Omega({\cal E},p) \|_{\infty}  \big)^m = (\mu_{\max})^m.
\end{eqnarray}

Since, the supremum in (\ref{normach}) is attained using a
pure state input and   (\ref{bndfin}) holds for all pure inputs 
$|\Psi\rangle$, we conclude that the upper bound
\begin{eqnarray}
\nu_p({\cal E}^{\otimes m}) \leqslant (\mu_{\max})^{m/p}
\label{upperbound}\;,
\end{eqnarray}
 holds for all  pairs of integers $m$ and $p$.
 
\subsection{Multiplicativity condition}\label{s:cond}

The bound (\ref{upperbound}) leads to a sufficient condition 
for multiplicativity.  We state this formally, and give a relate condition
as a corollary.
\begin{thm}   \label{thm:main}
The channel ${\cal E}$ has the multiplicativity property (\ref{normachmolt}) 
if  the largest singular value of  $\Omega({\cal E},p)$ satisfies
\begin{eqnarray}  \label{condb}
  \| \Omega({\cal E},p) \|_{\infty} = \big[\nu_p({\cal E})\big]^p 
\end{eqnarray}  
\end{thm}
\begin{cor}  \label{multcor}
The channel ${\cal E}$ has the multiplicativity property (\ref{normachmolt}) 
if the largest singular value 
   of $\Omega({\cal E},p)$  is also an eigenvalue of $\Omega({\cal E},p)$
   with a product eigenvector of the form  $|\phi\rangle^{\ot p}$ 
 \end{cor}
 
 To prove Theorem~\ref{thm:main}, observe that in the notation of
 the preceding section ({\ref{condb})
 can be written as $\mu_{\max} = \big[\nu_p({\cal E})\big]^p$
Then (\ref{upperbound}) implies
\begin{eqnarray}
\nu_p({\cal E}^{\otimes m}) \leqslant (\mu_{\max})^{m/p} = 
\big[ \nu_p({\cal E}) \big]^m .
\end{eqnarray}  On the other hand,
one always has
\begin{eqnarray*}
  \nu_p({\cal E}^{\otimes m}) \geqslant  
\| {\cal E}^{\otimes m}(\gamma_{\max}^{\otimes m}) \|_p
   = \|  {\cal E}(\gamma_{\max}) \|_p^m =  \big[ \nu_p({\cal E}) \big]^m
\end{eqnarray*}
where $\gamma_{max}$ denotes the state which achieves the supremum
for $\nu_p({\cal E})$.  Combining these inequalities gives
$\nu_p({\cal E}^{\otimes m}) = \big[ \nu_p({\cal E}) \big]^m$.   
\hskip1cm {\bf QED}

To prove the corollary, observe that its hypothesis
holds if and only if there is a state
 $ |\phi \rangle $ in ${\cal H}$ such that
\begin{eqnarray}  \label{eval}
\mu_{\max}   =
\langle\boldsymbol \Phi  |  
 \Omega({\cal E},p) \, |\boldsymbol{\Phi} \rangle
 =   \mbox{Tr}[{\cal E} \big(| \phi \rangle \langle \phi | \big)] ^p
\end{eqnarray}
where the second equality used (\ref{omegaprop}) and our convention that
$|\boldsymbol \Phi \rangle =  |\phi \rangle^{\ot p}$.   But it is always 
true that
 \begin{eqnarray} \label{corbnd}
    \mbox{Tr}[{\cal E} \big(| \phi \rangle \langle \phi | \big)] ^p
   \leqslant \sup_{\gamma}  \mbox{Tr}[{\cal E}(\gamma)]^p \equiv 
   \big[ \nu_p({\cal E}) \big]^p
   \end{eqnarray}
so that $\mu_{\max}   \leqslant  \big[ \nu_p({\cal E}) \big]^p$. Combining this
with~(\ref{upperbound}) when $m = 1$, implies that 
$\mu_{\max}  =  \big[ \nu_p({\cal E}) \big]^p$so that the hypothesis
of Theorem~\ref{thm:main} holds.   \hskip1cm {\bf QED}

In Section~\ref{s:qubits} we will see that the condition in 
Theorem~\ref{thm:main} is not necessary. 
There are unital qubit CPT maps, which are known to be multiplicative, 
but do not satisfy (\ref{condb}).
Verifying the hypothesis of Corollary~\ref{multcor} requires
that one find an eigenvector as well as the largest singular value of an
operator, but does not require knowledge of $\nu_p({\cal E})$; 
condition (\ref{condb}) does require the latter, 
but  does not require computation of any eigenvectors.
In general, (\ref{condb}) seems easier to check.  However, in the examples we
analyzed, both conditions hold and the process of verifying one
easily yields the other.  It would be interesting to know 
if~(\ref{condb}) implies
that the singular value of $\Omega({\cal E},p)$ is also an eigenvector
with a product eigenvalue as in Corollary~\ref{multcor}.

\section{Applications}\label{s:applications}

\subsection{Qubit channels}\label{s:qubits}

\subsubsection{Notation}

We illustrate our condition by looking at some examples
of qubit channels, for which will use notation similar to that 
introduced in \cite{KR2,RSW}.
Any $2 \times 2$ matrix can be represented in the basis consisting the 
$2\times 2$ identity matrix $\openone$ and the
three Pauli matrices which we often write as a formal vector 
$\vec{\sigma}\equiv(\sigma_1,\sigma_2,\sigma_3)$. 
In this basis a density matrix can be written as
$\gamma = \tfrac{1}{2} \big[ \openone + \vec{w} \cdot \vec{\sigma} \big]$
with $ \vec{w}$ in ${\bf R}^3$ and $| \vec{w}| \leqslant 1$.  The density
matrix is   pure  if and only if $| \vec{w}|= 1$.  
Any linear map $\Phi$
on a qubit, can be described by two real vectors 
$\vec{s}$, $\vec{t} \in {\bf R}^3$ and by   a
$3\times3$ real matrix $T$, through the expression
\begin{eqnarray}
\Phi(z_0 \openone + \vec{z}\cdot\vec{\sigma})
=(z_0 + \vec{s}\cdot \vec{z} ) \openone + (z_0 \vec{t} 
+ T \cdot \vec{z})\cdot  \vec{\sigma} 
\label{dualP}\;,
\end{eqnarray}
which holds for all $z_0\in \mathbb{C}$ and $\vec{z}\in \mathbb{C}^{3}$.
This corresponds to representing $\Phi$ in the basis
$\{ \openone, \vec{\sigma} \}$ by the $4 \times 4$ matrix
$\begin{pmatrix}  1 &  \vec{s}^{\, t} \\  \vec{t} & T \end{pmatrix}$
which we have written in block form (with the convention that
$\vec{t}$ corresponds to a column vector and $\vec{s}^{\, t}$ a row vector,
using the superscript $t$ to denote transpose).
It was shown in \cite{KR1} that it suffices to consider
$T$ diagonal with real elements $\{ \lambda_1,\lambda_2,\lambda_3\}$.
[In essence, a variant of the SVD (which leads to negative as well as
positive $\lambda_k$) can be applied to $T$ corresponding
to rotations on the input and output bases respectively.]

In this notation, $\Phi$ is trace preserving (TP) if and only if 
$\vec{s}=0$ and it is unital if and only if $\vec{t}=0$.    
Additional conditions under which the map is
positivity preserving or completely positive (CP) are  more complex.
A complete set of conditions for the map to be CPT was obtained in
\cite{RSW}. When $t_1 = t_2 = 0$, these CPT conditions reduce
to $(\lambda_1 \pm \lambda_2)^2 \leqslant (1 \pm \lambda_3)^2 - t_3^2$,
as shown in Refs. \cite{FA1,RSW}.
Since the dual map of $\Phi$ is represented by the adjoint matrix, 
it satisfies,
\begin{eqnarray}
\wh{\Phi}(z_0 \openone + \vec{z}\cdot\vec{\sigma})
=(z_0 + \vec{t}\cdot \vec{z} ) \openone + (z_0 \vec{s} 
+ T^{t} \cdot \vec{z})\cdot  \vec{\sigma} 
\label{dualqubit}.
\end{eqnarray}

Since ${\cal H}$ is now 2-dimensional, the left shift $L_2$ is 
simply the SWAP operator $S$  which satisfies
\begin{eqnarray}  \label{SWAP}
S =  \tfrac{1}{2}  \big[ \openone\otimes\openone + {\sigma}_1\otimes\sigma_1
+ {\sigma}_2\otimes\sigma_2+ {\sigma}_3\otimes\sigma_3 \big]
\end{eqnarray}
It is then straightforward to use (\ref{omega}) to show that
\begin{eqnarray}
\Omega(\Phi,2)  =  \wh{\Phi}^{\otimes 2}(S)  
   =  \tfrac{1}{2}  \Big[ \big(1+|\vec{t}|^2 \big)
\openone\otimes\openone + \sum_{j=1}^3 \lambda_j^2 {\sigma}_j\otimes\sigma_j  
  +\sum_{j=1}^3 \lambda_j t_j \big( \openone \otimes\sigma_j
+\sigma_j \otimes \openone \big) \Big]\;.
\label{uno}
\end{eqnarray}

\subsubsection{Unital maps}

For qubit channels the
conjecture (\ref{normach})
has been extensively studied in \cite{KR1,King1,King2,King5}.
Multiplicativity has been proven for all $p$ for unital qubit 
channels  \cite{King1}
and for $p=2$ for all qubit channels (Theorem ~2 of \cite{King1}).
Here we will use the case  $p=2$ 
to illustrate the multiplicativity-criterion presented
in Section~\ref{s:cond}.

It will be useful to choose the subscript ``max'' in $\{ 1,2,3 \}$ so that
$|\lambda_{\max}| = \max_{k} |\lambda_k| $.
For unital qubits maps, the maximum $\ell_2$-norm of 
$\Phi$ can be achieved with an input state  of the form
$\tfrac{1}{2} \big[ \openone \pm \sigma_{\max} \big]$ for which the
output  $\tfrac{1}{2}[ \openone \pm  \lambda_{\max} \sigma_{\max}]$
has eigenvalues $ \tfrac{1}{2}[ 1 \pm \lambda_{\max}]$ and
\begin{eqnarray}  \label{nu2unq}
 \nu_2(\Phi) = \tfrac{1}{\sqrt{2}}  \sqrt{1 + \lambda_{\max}^2 }
 \end{eqnarray} 
When $\Phi$ is unital, $\vec{t}=0$ and the third term in the expression
(\ref{uno}) vanishes.
It then follows that in the product basis
$\{ |00 \rangle, |01 \rangle, |10 \rangle , |11 \rangle  \}$,
the operator $\Omega({\Phi}, 2)$ is represented by the matrix
\begin{eqnarray}   \label{Omeg.un.mx}
 \frac{1}{2} \begin{pmatrix}     
1   + \lambda_3^2  & 0 & 0 & \lambda_1^2 - \lambda_2^2 \\
0 & 1  - \lambda_3^2 &   \lambda_1^2 + \lambda_2^2 & 0 \\
0 &  \lambda_1^2 + \lambda_2^2 &    1   - \lambda_3^2   & 0 \\
\lambda_1^2 - \lambda_2^2  & 0 & 0 &   1   + \lambda_3^2  
\end{pmatrix}~.  
\end{eqnarray}  
This is easily seen to have two non-zero $2 \times  2$ blocks.  
The ``inner'' block has eigenvalues 
$  \tfrac{1}{2} \Big[ 1 - \lambda_3^2  
\pm  \big( \lambda_1^2 + \lambda_2^2\big) \Big]$
with eigenvectors $\tfrac{1}{ \sqrt{2}}(0, 1, \pm 1, 0)^t$ 
corresponding to the Bell states  
$\tfrac{1}{ \sqrt{2}} (|01\rangle \pm |10\rangle) $.
The ``outer'' block has eigenvalues 
$ \tfrac{1}{2} \Big[  1 + \lambda_3^2  \pm  
\big( \lambda_1^2 - \lambda_2^2\big)  \Big]$
with eigenvectors $\tfrac{1}{ \sqrt{2}}(1, 0, 0, \pm 1)^t$ 
corresponding to the 
Bell states  $\tfrac{1}{ \sqrt{2}} (|00\rangle \pm |11\rangle) $.   Since
$\Omega({\Phi}, 2)$ is Hermitian, its singular values are simply the
absolute values of the eigenvalues above.

When the $|\lambda_k|$ are distinct for $k = 1,2,3$, the singular values of
 $\Omega({\Phi}, 2)$ are all distinct and correspond to maximally entangled,
 rather than product, states.   Moreover, one of the singular values is always
 strictly greater than $\nu_2(\Phi)$.    For example, when 
  $|\lambda_{\max}| = |\lambda_3|$,  one of the ``outer'' eigenvalues  
equals $\nu_2(\Phi)^2 + \tfrac{1}{2} \big( \lambda_1^2-\lambda_2^2\big) $
which is strictly greater than (\ref{nu2unq}) unless  
$|\lambda_1| = |\lambda_2|$.
Therefore, although $\Phi$ is multiplicative, 
it does not satisfy (\ref{condb}).
This establishes that (\ref{condb}) is not a necessary condition for
multiplicativity.
   
Now consider the case $\lambda_3 > \lambda_1 = \lambda_2 \geqslant 0$;
such channels are sometimes called ``two-Pauli'' channels  \cite{BFS}.  
The image of the Bloch sphere is an ellipsoid shaped like an American football.
For these channels, the  ``outer''  block in  (\ref{Omeg.un.mx}) 
is diagonal, its (degenerate)  eigenvalue  
$ \tfrac{1}{2} \big( 1 + \lambda_3^2 \big) = [\nu_2(\Phi)]^2$ is the 
largest singular value of  $\Omega({\Phi}, 2)$ and the corresponding
eigenvectors $|00 \rangle$ and $|11 \rangle$ are product states.
Thus, Theorem~\ref{thm:main} implies that 
the channel satisfies (\ref{normachmolt}).

\subsubsection{Non-unital maps}\label{s:nonunital}

We now consider channels similar to those above, but with
the  image ellipsoid shifted along the longest axis.
It suffices to consider 
$|\lambda_3| \geqslant \lambda_1 = \lambda_2 \geqslant 0$ 
and   $t_1=t_2=0$.     The same results hold for permutations of $ 1,2,3$ 
and for $\lambda_1 = \lambda_2 \leqslant 0$.
  However, the analysis in the basis we have chosen to represent
 $\Omega({\Phi}, 2)$ is simplest when $|\lambda_{\max}| = |\lambda_3|$.
The matrix representing  $\Omega({\Phi}, 2)$ is  
\begin{eqnarray}   \tfrac{1}{2} \begin{pmatrix}  \label{Omeg.t3}
    1  + (t_3    +   \lambda_3)^2  & 0 & 0 & 0 \\
      0 & 1   +   t_3^2  -   \lambda_3^2 &   2 \lambda_1^2  & 0 \\
      0 &  2 \lambda_1^2  &    1   +   t_3^2  -  \lambda_3^2   & 0 \\
      0  & 0 & 0 &   1  + (t_3    -   \lambda_3)^2
\end{pmatrix} \end{eqnarray}  
which has an ``inner'' block with eigenvalues
$ \tfrac{1}{2} \Big[ 1 + t_3^2 - \lambda_3^2  \pm  2 \lambda_1^2  \Big]$
 and a diagonal ``outer'' block with eigenvalues 
 $  \tfrac{1}{2}  \big[1 + (t_3 \pm \lambda_3)^2 \big]$
    and product eigenvectors. 
    One can verify that the largest singular value
 is $ \tfrac{1}{2}  \big[1 + (|t_3| + |\lambda_3|)^2 \big]$.
To see that this equals $[\nu_2(\Phi)]^2$, observe that 
 the optimal input state is 
$\tfrac{1}{2}[ \openone + \tfrac{t_3}{|t_3|} \sigma_3 \big]$ for
which the output state has eigenvalues 
$ \tfrac{1}{2}  \big[1 \pm  (|t_3| + |\lambda_3|)^2 \big]$.
Thus, we can again use Theorem~\ref{thm:main} to conclude
that (\ref{normachmolt}) holds.

The methods introduced here are able to handle qubit channels for which
the image of the Bloch sphere is an elongated ellipsoid with a symmetry axis,
i.e., in the shape of an American football, both when the channel is unital
and when it is shifted in the direction of the longest axis.   However, it 
can not handle these channels if the shift is orthogonal to the longest axis,
i.e., if $t_3 = 0$ but $t_2 \neq 0$ above.    When the ellipsoid has a symmetry
axis but $|\lambda_1| = |\lambda_2| \geqslant  |\lambda_3|$ 
so that it is shaped like a flying saucer, 
the methods used here can not prove multiplicativity.
Even for unital channels, for which multiplicativity has been established
\cite{King2}, neither of the conditions in Theorem~\ref{thm:main} holds.

\subsection{Shifted depolarizing channels}  \label{s:shft_dp}

\subsubsection{Shifting and generalizing the depolarizing channel}

The unital qubit  map with $\lambda_k  =  \pm |\lambda_{\max} |$ for all $k$, 
is a special case of the depolarizing channel which has the form
${\cal E}(\gamma) = (1-x) (\mbox{Tr} \gamma) \frac{1}{d} \openone + x \gamma$.
It is CPT for $- \frac{1}{3} \leqslant x \leqslant 1$.
The non-unital qubit map  which takes 
\begin{eqnarray}
\gamma =  \tfrac{1}{2} \big[ \openone + 
  \vec{w}\cdot\vec{\sigma}\big] & \mapsto & \tfrac{1}{2} \big[ \openone + 
  (\vec{t} + \lambda \vec{w}) \cdot\vec{\sigma} \big]   \\
     & = & (1 - |\vec{t}| - \lambda)  \tfrac{1}{2}   \openone +  |\vec{t}| 
        \tfrac{1}{2} \big[ \openone + \wh{t} \cdot\vec{\sigma} 
\big] + \lambda \gamma  \nonumber
  \end{eqnarray}
can be regarded as a shifted depolarizing channel because it shifts the
 output toward the point $ \wh{t} $ on the Bloch sphere.
By rotating coordinates so that $\vec{t} = (0,0,t_3)$, this is a special case
of the qubit maps considered in Section~\ref{s:nonunital} above.  
It is then natural to define a shifted depolarizing channel in dimension $d$ by
\begin{eqnarray}  \label{shft1dp}
  {\cal E}(\gamma) = a (\mbox{Tr} \gamma) \frac{1}{d} \openone +
      b (\mbox{Tr} \gamma) | \psi \rangle \langle \psi | + c \gamma
\end{eqnarray} 
with the state $ | \psi \rangle $ fixed and $a + b + c = 1$.   When
$a,b,c$ are positive, this channel is a convex combination of the identity map
and two completely noisy channels which maps all states to  
$\frac{1}{d} \openone$ and to  $| \psi \rangle \langle \psi |$,
respectively.

We now consider the more general class of channels of the form
\begin{eqnarray}  \label{shft_dp}
{\cal E}(\gamma) = (1-c) (\mbox{Tr} \gamma)  \rho       + c \gamma
  \end{eqnarray}
where $\rho$ is a fixed density matrix.  
For  $\rho = \frac{1}{d} \openone$, 
this is the usual depolarizing channel;  for
 $\rho = \frac{1}{a+b} \big[ \frac{a}{d} \openone + b | \psi 
\rangle \langle \psi | \big]$.
it is the shifted depolarizing channel  (\ref{shft1dp}).

When $c \geqslant 0$ additivity was proved for the 
depolarizing channel in $d$ dimensions
using a majorization argument  \cite{FA2} from which multiplicativity
immediately follows;  for   $-\tfrac{1}{d^2 - 1}  \leqslant c \leqslant 1$,
(which is the range for which the map is CPT) multiplicativity of the 
depolarizing channel in $d$ dimensions was proved  in \cite{King4}. 
Neither shifted depolarizing channels nor the generalization (\ref{shft_dp}) 
seem  to have been explicitly considered
 in the literature before.   One could obtain a proof  of
 multiplicativity for $p = 2$ when $c > 0$ by verifying that 
 the positive element
 condition in \cite{KR2} is satisfied.   (In fact, these maps satisfy the
 stronger condition considered in \cite{KNR}.)   However, neither of these
 positive element conditions can  be verified when $c < 0$.  
 By contrast, the method presented here can establish 
 multiplicativity when $p = 2$ for all CPT maps of the form (\ref{shft_dp}),
 including those with $c < 0$.

\subsubsection{Convex combinations of the identity and completely noisy maps} 

It will be useful to write the spectral decomposition of $\rho$
as $\rho = \sum_j a_j |j \rangle \langle j| $ with the
eigenvalues $a_j$ in decreasing order.    Then, for $c > 0$,
the state ${\cal E}(|1 \rangle \langle 1|)$ majorizes all  outputs so that 
$[\nu_p({\cal E})]^p = [ca_1 + (1-c)]^p +  c^p \sum_{j > 1}^d a_j^p $.

Since,  
$\wh{{\cal E}}(B) = (1-c) \big[ \mbox{Tr} B \rho \big] \openone + c B$,
we have
\begin{eqnarray}
   \wh{{\cal E}} \big( |j \rangle \langle k| \big) =    
(1-c)   \langle k|  \rho |j \rangle   \openone +
       c |j \rangle  \langle k|    
   =    (1-c)  \delta_{jk}  a_k     \openone +
       c |j \rangle  \langle k| ,
       \end{eqnarray} 
and
\begin{eqnarray}
   \Omega({\cal E},2)  & = & \big(\wh{{\cal E}} \ot \wh{{\cal E}} \big)(S)
       =  \sum_{jk} \wh{{\cal E}} \big( | j \rangle \langle k|  \big) \ot  
\wh{{\cal E}} \big( |k \rangle \langle j|  \big)     \nonumber \\
      & = &  \sum_{jk} \bigg[ (1-c)^2 \delta_{jk} a_k^2 \openone \ot 
\openone +  c(1-c)
         \delta_{jk} a_k \Big( \openone \ot |k \rangle \langle k| + 
|k \rangle \langle k| \ot \openone \Big)
          + c^2 \;|j \rangle \langle k | \ot |k \rangle \langle j| \bigg] \nonumber \\
          & = & (1-c)^2  \big( \mbox{Tr}  \rho^2 \big)  
\openone \ot \openone  + c(1-c)
          \big[  \openone \ot \rho  + \rho \ot \openone  \big] + c^2 S.
       \end{eqnarray}
From this it is easy to see  that    $ \Omega({\cal E},2)$ has 
$d$ product eigenvectors of the form $|kk \rangle$ with 
 eigenvalues 
 \begin{eqnarray}(1-c)^2 \big( \mbox{Tr}  \rho^2 \big)  + 2c(1-c) a_k + c^2 =
     [(1-c)a_k + c]^2 + (1-c)^2 \sum_{j \neq k} a_j^2, 
     \end{eqnarray}   and
$\tbinom{d}{2}$ blocks  of the form  
$\big[(1-c)^2 \big( \mbox{Tr}  \rho^2 \big) + 
c(1-c)(a_j + a_k) \big] \openone_2 + c^2 \sigma_x$,
 with      eigenvalues 
\begin{eqnarray}
(1-c)^2 \big( \mbox{Tr}  \rho^2 \big)  + c(1-c) (a_j + a_k) \pm  c^2
\end{eqnarray}
and entangled eigenvectors 
$2^{-1/2} \big( |jk \rangle \pm |kj \rangle \big)$.
When $c > 0$ all eigenvalues are non-negative and the largest singular value
is  $[(1-c)a_1 + c]^2 + (1-c)^2 \sum_{j  > 1} a_j^2 =  [\nu_2({\cal E})]^2$ 
associated with the product eigenvector $|11 \rangle$.
Therefore, one can use Theorem~\ref{thm:main}, or Corollary~\ref{multcor},
to conclude that the channel (\ref{shft_dp}) 
is multiplicative for $p = 2$ when $c > 0$.
 
 \subsubsection{CPT Maps with a negative contribution from the identity} 
   
To analyze the case  $c < 0$,  write $c = -x$ with $x = |c| > 0$, and recall
that we assumed that the $\{ a_j \}$ are decreasing.  
It can still happen that all eigenvalues of $ \Omega({\cal E},2)$ 
are non-negative,
in which case the largest singular value is   
$[(1+x)a_d -x]^2 + (1+x)^2 \sum_{j  < d} a_j^2$
associated with the product eigenvector $|dd \rangle$.  It turns out that 
the requirement  that ${\cal E}$ be CPT suffices to ensure that the 
eigenvalues of $ \Omega({\cal E},2)$ are non-negative.  Therefore,
any CPT map of the form (\ref{shft_dp}) is multiplicative for $p = 2$.

To see the relevance of the CPT condition, observe that the CP requirement
that $({\cal E} \ot \openone) \Big( \sum_{jk} |j \rangle \langle k| 
\ot |j \rangle \langle k| \Big)$ 
(which is the Choi matrix)  is positive semi-definite holds if and only if 
$B = (1+x) \rho - x \begin{pmatrix}  1 & \cdots & 1 \\
\vdots & ~  & \vdots \\ 1 & \cdots & 1 \end{pmatrix}$ 
is positive semi-definite.
Then $B$ has non-negative diagonal elements,  which gives
    \begin{eqnarray} \label{xbnd}
       (1+x)a_j - x   \geqslant 0  ~~ \Rightarrow   ~~     
       x   \leqslant  \tfrac{a_j}{1-a_j} &&  
       ~~ \Rightarrow   ~~  \tfrac{x}{1+x}   \leqslant a_j.
    \end{eqnarray}
All $2 \times 2$ principle minors of $B$ are non-negative,  which implies
  \begin{eqnarray}
      (1+x)^2 a_1 a_2 - x(1+x) (a_1 + a_2) \geqslant 0
   \end{eqnarray} 
Now, all eigenvalues of   $ \Omega({\cal E},2)$  will be positive if
$(1-c)^2 \big( \mbox{Tr}  \rho^2 \big)  + c(1-c) (a_j + a_k) -  
c^2 \geqslant 0$     
for all $j,k$.  But the most negative of these is
 \begin{eqnarray}
      (1+x)^2 \big( \mbox{Tr}  \rho^2 \big)  -x(1+x) (a_1 + a_2) -  x^2 
      & \geqslant &
      (1+x)^2 (a_1^2 + a_2^2) - (1+x)^2 a_1 a_2  -x^2    \nonumber \\
      & = & (1+x)^2 \Big[ a_1^2 + a_2^2 -   a_1 a_2 - 
      \big( \tfrac{x}{1+x} \big)^2 \Big] \\
      & \geqslant & (1+x)^2  \big[ a_1^2 + a_2^2 - 2a_1 a_2 \big] = 
     (1+x)^2 (a_1 - a_2)^2 \geqslant 0,  \nonumber
  \end{eqnarray}
where the second inequality used (\ref{xbnd}) with $j = 1, 2$ 
to conclude that $\big(\tfrac{x}{1+x} \big)^2   \leqslant a_1 a_2$.
           
% \pagebreak
 
\subsection{The Werner-Holevo channel}\label{s:werner}

In our final example, we apply our condition for $p = 3,4$ as well as $p = 2$.
We study the channels ${\cal W}_d$ introduced in \cite{HW}
to show that multiplicativity does not hold for sufficiently large $p$.
The channel ${\cal W}_d$ is defined on a $d$ dimensional Hilbert space as
\begin{eqnarray}
{\cal W}_d(\gamma) &\equiv& \frac{1}{d-1} \big[ (\hbox{Tr} \, \gamma )\, \openone_d - \gamma^T \big]
=  \frac{1}{d-1} \sum_{j < k} W_{jk}^{\dag} \gamma \, W_{jk}
\label{werner}
\end{eqnarray}
with $\openone_d$ the identity operator on $\cal H$, $\gamma^T$ 
the matrix transpose with respect to some fixed basis $\{|i\rangle\}$,
and $W_{jk}$ the anti-Hermitian operator 
$| j \rangle\langle k| -  |k \rangle\langle j|$.  (We will often suppress the
subscript $d$ and simply write  ${\cal W}$ for ${\cal W}_d$.)  As observed in
\cite{HW}, any pure input state yields an output state 
${\cal W}(|\psi \rangle \langle \psi |)$
with eigenvalues ${1}/{(d-1)}$ with multiplicity $d-1$.
 This implies
\begin{eqnarray}  \label{nupW}
   \nu_p({\cal W}_d) = (d-1)^{(1-p)/p}
\end{eqnarray}
Werner and Holevo showed that for $d=3$ and $p>4.79$ this map is not
${\ell}_p$ multiplicative, by showing that maximally entangled inputs  yield
output ${\ell}_p$ norm greater than  $(d-1)^{(1-p)/p}$.  For $d > 3$, they
also showed that multiplicativity fails for sufficiently large $p$.
Although their results strongly suggest that multiplicativity does hold
for smaller $p$, they do not preclude the possibility that it fails
 with inputs that are partially entangled.    Our results show that this
 cannot happen when $p = 2,3,4$ and $d \geqslant 2^{p-1}$.

 The multiplicativity of $\cal W$ for $p=2$ was 
established in \cite{KR2}; the additivity of minimal output entropy and
Holevo capacity was proved in
\cite{MY} and \cite{DHS}; and, recently, a short elegant proof of 
 multiplicativity for all $1 \leqslant p \leqslant 2$ was given in \cite{AF}.
Here we use Theorem~\ref{thm:main} to give
another proof of (\ref{normachmolt})   for $p=2$, and then
consider multiplicativity of $\Omega({\cal W},p) $  for integer $p > 2$.  

For $p = 2$ it is straightforward to show that (or see Appendix ~\ref{s:WH1})
\begin{eqnarray}
\Omega({\cal W},2) = ({\cal W} \ot {\cal W})(S) = 
 \frac{1}{(d-1)^2} \big[(d-2)  \openone\otimes \openone + S \big] .
\label{teta}
\end{eqnarray}
with $S$ the SWAP on ${\cal H}\otimes{\cal H}$. 
The eigenvalues of $\Omega({\cal W},2) $ can be computed 
from those of $S$  which has a diagonal block with $d$ product states 
$|jj \rangle$ as eigenvectors with eigenvalue $1$,
and $\binom{d}{2}$ blocks of the
form $\begin{pmatrix}  0 & 1 \\ 1 & 0 \end{pmatrix}$ with eigenvalues
$+1$ and $-1$ corresponding to the entangled states
$\frac{1}{\sqrt{2}} \big( |jk \rangle \pm |kj \rangle \big)$.   This yields
eigenvalues $\frac{1}{d-1}$ with multiplicity $\frac{d(d+1)}{2}$ and 
$\frac{d-3}{d-1}$ with multiplicity $ \frac{d(d-1)}{2}$.  
For $d \geqslant 3$, these are also the
singular values of $ \Omega({\cal W},2)$; for $d = 2$,   $\frac{1}{d-1}$ is the
only singular value.  In both cases 
$\|  \Omega({\cal W},2) \|_{\infty} = \frac{1}{d-1} =  \nu_2({\cal W})$.
Therefore, (\ref{condb}) 
is satisfied and the result follows from Theorem~\ref{thm:main}.

To study $p > 2$, we first observe that (\ref{WHgen}) implies that
 $\Omega({\cal W},p) $ is a linear combination of permutation
matrices.  This has some important consequences.
\begin{itemize}
 
 \item[a)]   $\Omega({\cal W},p) $ has a large number of invariant subspaces,
giving it a block diagonal structure.    Each block describes the restriction
of $\Omega({\cal W},p) $ to a subspace spanned by all permutations of
a vector
$| \xi_{k_1} \xi_{k_2}  \ldots \xi_{k_p} \ket$ with indices
$k_1 \leq k_2 \leq \ldots \leq k_p$.   

 \item[b)]  All  row and column sums are equal.   Moreover, 
(\ref{WHrowsum}) implies that every row and
column sum of $\Omega({\cal W},p) $ or, equivalently,  
  of each block, is exactly $(d-1)^{1-p}$,
which is also the value of  $[\nu_p({\cal W})]^p$.

 \end{itemize}
 It follows immediately from (b) that $(d-1)^{1-p}$ is an eigenvalue  of
 each block of $\Omega({\cal W},p) $ 
 and, hence, an eigenvalue of $\Omega({\cal W},p) $ with 
 very high degeneracy.   Therefore,  $\Omega({\cal W},p) $
 can have a singular value greater that $[\nu_p({\cal W})]^p$
 only if some block has a singular value greater than $(d-1)^{1-p}$.
 The following lemma, which is proved in Appendix~\ref{app:bigblk},
shows that it will suffice to consider this question for one of the largest blocks.
\begin{lemma}  \label{bigblk}
When $d \geq p$, the largest singular value of $\Omega({\cal W},p) $
is a singular value of each of the $p! \times p!$ blocks  \linebreak representing
the restriction
of $\Omega({\cal W},p) $ to a subspace of $(\mathbb{C}^d)^{\ot p}$ 
spanned by all permutations of a vector $| \xi_{k_1} \xi_{k_2}  \ldots \xi_{k_p} \ket$
with distinct $k_p$.
\end{lemma}

Based on this and the structure of the largest blocks as described
in Appendx~\ref{app:irrep}, we make the following
\begin{conj}   
The $\ell_p$ multiplicativity
relation (\ref{normachmolt}) holds
 for the channel ${\cal W}_d$  when the dimension
$d \geqslant  2^{p-1}$.
\end{conj}
This conjecture is proved for $p = 2,3,4$.   For larger $p$ we have shown
in Appendix~\ref{app:irrep} that  the largest block of $\Omega({\cal W},p) $ 
has two eigenvectors which
transform as the two one-dimensional representations of ${\cal S}_p$.  
The corresponding eigenvalues are $(d-1)^{1-p}$ and $(d-1)^{-p} (d - 2^p +1)$.
When $d \geq  2^{p-1}$, $|d - 2^p +1|  \leq  d-1$.   
Moreover, no other singular values
have the symmetry associated with a one-dimensional  representation of  
${\cal S}_p$. Thus, if we knew that the
largest singular value of $\Omega({\cal W},p) $  must be associated
with a one-dimensional irreducible representation, we could conclude
that the largest singular value of $\Omega({\cal W},p) $ is $d-1$,
proving the conjecture.

Now we consider $p = 3, 4$.  The results in 
Appendix~\ref{s:WH1} can be used to write $\Omega({\cal W},p) $ 
explicity as
\begin{eqnarray}
 \Omega({\cal W},3) & = & \tfrac{1}{(d-1)^3} \Big[ (d-3) \openone + 
  \sum_{a < b} S_{ab}
  - R_3 \Big] \;, \label{omega3}\\
   \Omega({\cal W},4) & = & \tfrac{1}{(d-1)^4} \Big[ (d-4) \openone
 + \sum_{a < b} S_{ab}  
   - \sum_{a < b < c}  R_3(a,b,c)  + R_4 \Big] \label{omega4} \;, 
   \end{eqnarray}
where the shift $R_3(a,b,c)$ is defined in Appendix~\ref{s:shift}.
The block structure 
of $\Omega({\cal W},p) $ for $p = 3,4$ is summarized in 
Table~\ref{WHtable}.
In this table, $i,j,k, \ell$ always denote distinct indices.
For readability, 
$(d-1)^p \mu_{\max}$ is reported in the last three columns,   
and should be compared to $(d-1)^p [\nu_p({\cal W})]^p = (d-1)$.
\begin{table}[h] 
\begin{tabular}  {|c|c|c|c|ccc|}   \hline \hline
   number  & size & type of & non-neg   & 
\multicolumn{3}{c|} {max sing value $\times (d-1)^p$ } \\
    of blocks & & vectors  &  elements  & ~$d = 3$ ~ & ~  $d = 4$ ~ & ~ 
   $d \geqslant 5 $ ~ \\ \hline
   \multicolumn{7}{|c|}{ $p=3$}     \\ \hline
   % $[(d-1)\nu_3]^3  =  d-1$
      $ d   $  &    $ 1 \times 1  $ &  $    |kkk \rangle  $  & yes & 
$2$ & $3$ & $d-1$\\
     $ d(d-1) $   &   $  3 \times 3 $  & $  |jjk \rangle $  & yes  & 
$2$ & $3$ & $d-1$ \\
   $  \tbinom{d}{3}    $   &   $   6 \times 6    $   &   $   
|ijk \rangle $ & no& 
$ 4 $  & $3$ & 
   $\max \{d-1, \,  |7-d| \}$\\  \hline 
  %       \multicolumn{3}{c} ~~ \\ \hline  \hline
  \multicolumn{7}{|c|}{ $p=4$} \\ \hline
   $ d   $  &    $ 1 \times 1  $ &  $     |kkk k \rangle  
$ & yes & $2$ & $3$ & $d-1$ \\
   $ d(d-1) $   &   $ 4 \times 4  $   &   $  |jjjk \rangle $ & yes & $2$ 
& $3$ & $d-1$  \\
  $  \tbinom{d}{2}   $   &   $  6 \times 6  $   &   $  |jjkk \rangle  $
 & yes &  $2$ & $3$  & $d-1$ \\
 $   \tfrac{1}{2} d(d-1)(d-2) $   &   $  12 \times 12 $   &   $    
|ijkk \rangle $ & no & $ \sqrt{18} $ & $\sqrt{13}$ & 
 $\max\{ d-1, \sqrt{d^2-12d+45} \} $   \\
   $  \tbinom{d}{4}   $   &   $  ~ 24 \times 24 ~ $   &   $~~  
|ijk \ell \rangle ~~$ & no &  ~ &  11 & 
   $\max\{ d-1, |15-d| \} $\\ \hline \hline  
 \end{tabular} 
\caption{Block structure of $\Omega({\cal W},p)$. \label{WHtable}} \end{table}
For  $\Omega({\cal W},3)$ and $ \Omega({\cal W},4)$, all singular values
can be found explicitly with the help of  Mathematica, with the largest for\
each block shown in  Table~\ref{WHtable}.   The multiplicativity 
condition (\ref{condb}) holds if the largest singular value is $(d-1)^{1-p}$.
For $p = 3$, this holds for $d \geqslant 4$;
for $p = 4$, it holds for $d \geqslant 8$.  
For $p = 3$, an analytic argument, which does not require  determining the
eigenvalues of $\Omega({\cal W},3)$, is presented in Appendix~\ref{s:p3}.

\pagebreak
   
\section{Conclusion}\label{s:conclusion}
We have extended the method introduced in \cite{GIOVA1,GIOVA2} to study the 
maximal ${\ell}_p$-norms of a CPT map when $p$ is a fixed integer.  
 This yields a sufficient condition for  multiplicativity which requires
 only that one find the singular values of a particular matrix, rather than
 performing a full optimization.   Although the matrix will be 
 $d^p \times d^p$, it often has a block 
 structure which makes the problems quite tractable,
 as shown in several examples.   The condition is not necessary, but does
 allow us to prove new results about multiplicativity in several 
 interesting cases, 
 as well  as providing alternative proofs of known results.
 
\bigskip

\appendix

\section{Some operator properties}  \label{app}

\subsection{Hilbert-Schmidt duality}  \label{s:dual}

For a Hilbert space ${\cal H}$ the subspace of operators 
satisfying $\mbox{Tr} A^{\dagger} A < \infty$ also forms a
Hilbert space (the space of Hilbert Schmidt operators)
with respect to the inner product
\begin{eqnarray} \label{HSip}
  \langle A , B \rangle = \mbox{Tr} A^{\dagger} B
\end{eqnarray}
An operator (sometimes referred to as a ``superoperator'')
${\cal E}$ acting on this space has an adjoint which we
will denote $\wh{\cal E}$ and which satisfies
\begin{eqnarray}   \label{dualdef}
   \mbox{Tr}  [{\cal E}(A)]^{\dagger} B  =   
   \mbox{Tr} A^{\dagger} \wh{\cal E}(B)  ~~~
\forall ~A,B.
\end{eqnarray}
Because $[{\cal E}(A)]^{\dagger} = {\cal E}(A^{\dagger})$, by writing $C$ for 
$A^{\dagger}$ one easily sees that (\ref{dualdef}) is equivalent to
to the condition
\begin{eqnarray}   \label{dualdef2}
   \mbox{Tr}  [{\cal E}(C)]  B  =   \mbox{Tr} C \wh{\cal E}(B)  
   ~~~\forall ~B,C.
\end{eqnarray}
The map $\wh{\cal E}$ is often called the dual of $\cal E$ because
it is defined by the duality property of the Riesz  representation
theorem applied to the inner product (\ref{HSip}).
When ${\cal E}$ is a CPT map of the form 
(\ref{mappa}), its dual is the unital CP map with the form
\begin{eqnarray}
\wh{\cal E}(\gamma) = \sum_k A_k^{\dag} B A_k .
\label{dual1} \end{eqnarray}
One can verify, either directly from (\ref{HSip}) or by using (\ref{dual1}), 
that the
dual of the map ${\cal E}^{\otimes m}$ is given by the $m$-fold tensor product
of  the dual map of $\cal E$, i.e.
$\wh{{ \cal E}^{\otimes m}} = \big(\,\wh{{\cal E}}\,\big)^{\otimes m} $.

\subsection{Shift operators} \label{s:shift}

The shift operators defined in (\ref{shft}) are unitary and satisfy
$L_p  R_p = \openone$ so that $L_p ^{\dagger} = L_p ^{-1} = R_p$.
Moreover, if a vector $| \Psi \rangle$ in ${\cal H}^{\otimes p}$
has the expansion
\begin{eqnarray}
   | \Psi \rangle = \sum_{j_1 j_2 \cdots j_p}  c_{j_1 j_2 \cdots j_p}
     | \xi_{j_1} \xi_{j_2} \cdots  \xi_{j_p} \rangle
\end{eqnarray}
then 
\begin{eqnarray}
   L_p  | \Psi \rangle & = &  \sum_{j_1 j_2 \cdots j_p}  c_{j_1 j_2 \cdots j_p}
    | \xi_{j_2} \xi_{j_3} \cdots \xi_p  \xi_{j_1} \rangle \\
    & = &  \sum_{j_1 j_2 \cdots j_p}  c_{j_p j_1 \cdots j_{p-1}}
    | \xi_{j_1} \xi_{j_2} \cdots  \xi_{j_p} \rangle 
  \end{eqnarray}
so that $L_p $ induces a right shift on the expansion coefficients.
From this, it follows that, that $L_p $ and $R_p$ induce
left and right shifts on all product states, e.g.,
\begin{subequations}  \label{shift}   \begin{eqnarray}
L_p \; |\phi_1 , \phi_2, 
\cdots ,\phi_p\rangle &=& |\phi_2 , \phi_3, 
\cdots ,\phi_p, \phi_1\rangle  \label{shift.a} \\
R_p \; |\phi_1 , \phi_2, 
\cdots ,\phi_p\rangle &=& |\phi_p , \phi_1, 
\cdots , \phi_{p-1}\rangle \label{shift.b} 
\end{eqnarray}  \end{subequations} 
where $|\phi_1 , \phi_2,  \cdots ,\phi_p\rangle$ denotes
$|\phi_1 \rangle \otimes | \phi_2 \rangle \otimes   
\cdots \otimes | \phi_p \rangle$.
It also follows from (\ref{shift}) that the shift operators are independent
of the choice of orthonormal basis in (\ref{shft}).

To compute operators associated with the WH-channel, it will be
useful to observe that
\begin{subequations}  \label{shftrep} \begin{eqnarray}
     L_p  =   \sum_{m_1 \cdots m_p}
      | m_2 \cdots  m_p  m_1 \rangle \langle m_1 m_2 \cdots  m_p | ~~ ~\\
      R_p  =  \sum_{m_1 \cdots m_p} 
       |m_p  m_1 \cdots  m_{p-1}  \rangle \langle m_1 m_2 \cdots m_p | ~
\end{eqnarray}  \end{subequations}
where $| m_j \rangle$ denotes any orthonormal basis of ${\cal H}$.
It will also be useful to introduce some notation for shift operators
on a subset of ${\cal H}^{\otimes p}$.  For example,
write ${\cal H}^{\otimes 4} = 
{\cal H}_a \otimes {\cal H}_b \otimes {\cal H}_c \otimes {\cal H}_d$.
Then $L_3(a,b,d) $ denotes the operator which acts as a left
shift on ${\cal H}_a \otimes {\cal H}_b \otimes {\cal H}_d$ and the
identity on  ${\cal H}_c$, i.e.,
 \begin{eqnarray}
  L_3(a,b,d) = \sum_{m_1 \cdots m_4} 
     | m_2 \, m_4  \, m_3 \, m_1 \rangle \langle m_1 \, m_2 \, m_3 \, m_4 |.
     \end{eqnarray}
 The SWAP operators $L_2(a,b) = R_2(a,b)$ play such a 
 special role that we denote them as $S_{ab}$.   Using the
 standard method for writing any permutation as a product of
 cycles, one can see that any shift can be written as a product
 of SWAP operators, e.g. 
  $L_3(a,b,d) = S_{ab} S_{ad}$ and 
 $L_4(a,b,c,d) = S_{ab} S_{ac} S_{ad}$.

 \subsection{Tensor products of shifts} \label{s:shiftmp}
 
 When the underlying Hilbert is itself a tensor product ${\cal H}^{\otimes m}$,
 we will let $\mathbb{L}_p$ denotes the shift operator acting on $p$ copies
 of ${\cal H}^{\otimes m}$, e.g.,  
 $\mathbb{L}_3 | x, y, z \rangle = |y, z, x \rangle$
 with $x, y, z$ denoting vectors in ${\cal H}^{\otimes m}$.  Then,  
    $ \mathbb{L}_p = L_p^{\otimes m} = \big(L_{mp}\big)^m$.
To avoid notation with double subscripts, 
we prove this in the case $p = 3$.  Then
 \begin{eqnarray}
     \mathbb{L}_3 | x, y, z \rangle  & = &  \mathbb{L}_3   \nonumber
        | x_1,  x_2,  \cdots x_m, y_1, y_2, \cdots y_m, z_1, z_2, \cdots z_m 
        \rangle \\  \label{Lmpeq}
         & = &  | y_1, y_2, \cdots y_m, z_1, z_2, \cdots z_m, x_1,  x_2,  
        \cdots x_m \rangle \\
          & = &  L_3^{\otimes m}   | x_1,  x_2,  \cdots x_m, y_1, y_2, 
        \cdots y_m, z_1, z_2, 
        \cdots z_m \rangle
          \nonumber 
 \end{eqnarray}
 where the last line follows by writing
 \begin{eqnarray*}
    L_3^{\otimes m} = (L_3 \ot \openone \ot \cdots \ot \openone) 
         ( \openone \ot L_3 \ot \openone \ot \cdots \ot \openone) \cdots
         ( \openone \ot \openone \ot \openone \ot \cdots \ot  
          \openone  \ot L_3) 
  \end{eqnarray*}
  and observing that
  \begin{eqnarray*}
    (L_3 \ot \openone \ot \cdots \ot \openone) | x_1,  x_2,  
     \cdots x_m, y_1, y_2, \cdots y_m, z_1, z_2, \cdots z_m \rangle
      = | y_1,  x_2,  \cdots x_m, z_1, y_2, \cdots y_m, x_1, z_2, 
     \cdots z_m \rangle.
  \end{eqnarray*}
Note that it is also evident from (\ref{Lmpeq}) that  
$L_p^{\otimes m} = \big(L_{mp}\big)^m$.

\subsection{An important trace identity}  \label{s:trace}

We now show that for any set of operators $\{ B_1, B_2, \cdots B_p \}$ acting
on ${\cal H}$, 
\begin{eqnarray}
  \mbox{Tr}_{\cal H}[B_1 \; B_2 \; B_3 \cdots B_p]  
  =   \mbox{Tr}_{{\cal H}^{\otimes p}}
\,  \big[ B_1 \otimes B_2 \otimes \cdots \otimes B_p \big] L_p  \;, 
\label{lemma}  
\end{eqnarray} 
where we have introduced subscripts to emphasize that
 the trace in the left-hand side of Eq.~(\ref{lemma})
is performed on  $\cal H$, while the trace in the right-hand side is performed
on ${\cal H}^{\otimes p}$. To verify (\ref{lemma}) observe that
\begin{eqnarray*}
 \mbox{Tr}[B_1 \; B_2   \cdots B_p]   &=& \sum_{\xi_1}
\langle \xi_1 | B_1 \; B_2   \cdots B_{p-1} B_p | \xi_1 \rangle \\
 & = &  \sum_{\xi_1,\cdots,\xi_p}
\langle \xi_1 | B_1 |\xi_2 \rangle\langle \xi_2| 
 B_2  | \xi_3 \rangle \cdots  \langle \xi_{p-1} |B_{p-1} |\xi_p\rangle 
\langle \xi_p | B_p | \xi_1 \rangle
\nonumber\\ &=& \sum_{\xi_1,\cdots,\xi_p} 
\langle \xi_1, \xi_2, \cdots, \xi_p | B_1 \otimes B_2 \otimes \cdots
\otimes  B_p |\xi_2, \cdots \xi_p,\xi_1\rangle 
\\ & = & \mbox{Tr} \,  \big[ B_1 \otimes B_2 \otimes 
\cdots \otimes B_p \big]  L_p 
\;,
\end{eqnarray*} 
where a resolution of the identity operator $\openone$ of $\cal H$ 
was inserted between the products $B_j B_{j+1}$. 

\subsection{General permutations} \label{s:perm}

 Shifts are special cases of permutation operators.  Let
 $\Pi_p$ denote a permutation of $\{ 1, 2 \cdots p \}$ and
 ${\cal S}_p$ the set of all such permutations.
   We will
 write $\Pi(j)  = k_j$ for the permutation that takes $j \mapsto k_j$.
 For example, $L_p(j) = j+1$.   One can then define a permutation operator
 on  ${\cal H}^{\ot p}$ by
 \begin{eqnarray}  \label{permH}
 \Pi_p  \, | \xi_{j_1} \xi_{j_2} \cdots  \xi_{j_p} \rangle = 
      | \xi_{\Pi(j_1)} \xi_{\Pi(j_2)} \cdots  \xi_{\Pi(j_p)} \rangle 
  \end{eqnarray}
with $\{ |\xi_j \rangle \}$ an orthonormal basis for   ${\cal H}$
 as in (\ref{shft}). A permutation of the indices  $\{ 1, 2 \cdots p \}$
 induces a permutation on the $d^p$ product basis vectors ${\cal H}^{\ot p}$
 via (\ref{permH}).   Although we abuse notation by using the same letter 
 for both, there should  be no confusion.    The permutation operator on
 ${\cal H}^{\ot p}$ is  represented by a $d^p \times d^p$ matrix which has
 precisely one $1$ and $d^p - 1 ~~0$'s in each row and column.
 
The permutation which takes $k_1 \mapsto  k_2 \mapsto 
\cdots \mapsto k_q \mapsto k_1$
is called a cycle and written  $P = (k_1, k_2 \cdots k_q)$, i.e., 
$P(k_j) = (k_{j+1})$
 with the understanding that $P(k_q) = k_1$ and $\Pi(j) = j$ if $j$ 
does not appear as 
one of the $k_i$ in the cycle.  Any permutation can be written uniquely 
 as a product of {\em disjoint} cycles, and the length of the disjoint
 cycles in $\Pi P \Pi^{\dag}$ are the same as those in $P$.   For example
 $(13) L_5 (13) = (14532)$.     If a permutation  of $\{ 1, 2 \cdots p \}$ has
 a cycle decomposition with cycles whose length is strictly less than $p$,
 then some subset of $\{ 1, 2 \cdots p \}$ is invariant.     
 A permutation $\Pi_p$ whose
 shortest cycle is of length $p$ has no invariant subsets.  Permutations
 satisfying this condition, which is equivalent
 to   $(\Pi_p)^s(j) \neq j$ for $s< p$ and $(\Pi_p)^p(j) = j$ for all $j$, 
 are of particular interest. 
   
In fact, when all operators $B_i  = B$ are identical,
(\ref{lemma}) can be extended to any permutation
$\Pi_p$ of $ \{ 1, 2 \cdots p \} $ whose
  shortest cycle is length $p$.   One finds
  \begin{eqnarray}
  \mbox{Tr}_{\cal H} B^p   
 & = &  \sum_{\xi_1,\cdots,\xi_p}
\langle \xi_1 | B  |\xi_{\Pi_p(1)}  \rangle \langle \xi_{\Pi_p(1)} | 
 B  | \xi_{\Pi_p^2(1)} \rangle  \cdots     \langle \xi_{(\Pi_p)^p(1)} | B  
| \xi_1 \rangle
   \nonumber   \\ &=& 
  \sum_{\xi_1,\cdots,\xi_p} 
\langle \xi_1 | B  |\xi_{\Pi_p(1)}  \rangle \langle \xi_2| 
 B  | \xi_{\Pi_p (2)} \rangle  \cdots      \langle \xi_p | B  | \xi_{\Pi_p(p)} 
\rangle   \nonumber \\ 
& = &   \sum_{\xi_1,\cdots,\xi_p} 
\langle \xi_1, \xi_2, \cdots, \xi_p | B  \otimes B  \otimes \cdots
\otimes  B  |\xi_{k_1}, \xi_{k_2} \cdots \xi_{k_p} \rangle \nonumber  \\
 & = & \mbox{Tr}_{{\cal H}^{\ot p}} \,  \big[ B  \otimes B  
\otimes \cdots \otimes B  \big]  \Pi_p 
 =   \mbox{Tr}_{{\cal H}^{\ot p}}  B^{\otimes p} \Pi_p
\label{lemma1}\;,
\end{eqnarray} 
To see where the invariance condition is used, consider the
permutation $(153)(24)$.  Attempting to apply the process above yields
\begin{eqnarray*}
\mbox{Tr}_{\cal H} B^5  & = &   \sum_{\xi_1,\xi_3, \xi_5} \langle
\xi_1 | B  |\xi_5  
\rangle \langle \xi_5 | B  |\xi_3  \rangle   \langle \xi_3 | B^3  |\xi_1  
\rangle \\
& = &     \sum_{\xi_1,\xi_3, \xi_5}  
\langle  \xi_1,  \xi_5 , \xi_3 | B  \otimes B  
\otimes B^3  | \xi_5, \xi_3, \xi_1 
\rangle \\
& = &   \mbox{Tr}_{{\cal H}_a \ot {\cal H}_c \ot {\cal H}_e }[B 
\otimes B \otimes B^3]    L_3
\end{eqnarray*} 
or $\mbox{Tr}_{\cal H} B^5  = \mbox{Tr}_{{\cal H}^{\ot 3} }
[B \otimes B^3 \otimes B] L_3$
or  $\mbox{Tr}_{\cal H} B^5  = \mbox{Tr}_{{\cal H}^{\ot 2} }
[B \otimes B^4]    L_2$.

 \subsection{Double stochastic matrices} \label{s:stoch}
 
A double stochastic matrix \cite{HJ1} is a matrix with non-negative  
elements whose row and column
sums are all $1$, i.e., $B$ is double stochastic 
if and only if $b_{jk} \geqslant 0 ~~ \forall ~ j,k$
and
$\sum_j b_{jk} = \sum_k b_{jk} = 1$.  The vector $( 1, 1, \cdots 1)$ is always
an eigenvector with eigenvalue $1$.    Moreover, all other eigenvalues
satisfy $|\lambda_j| \leqslant 1$.    A permutation of $\{ 1, 2 \cdots p \}$ 
can be
represented by a matrix which has
 precisely one $1$ and $p - 1 ~~0$'s in each row and column.
This is a special type of   double stochastic matrix called a 
``permutation matrix''.     
Moreover, a permutation $\Pi_p$ of  $\{ 1,2 \cdots p \}$ has no
non-trivial invariant subspaces if and only if  its permutation matrix is 
indecomposable.
Note that the corresponding permutation operator on ${\cal H}^p$, represented
by  a $d^p \times d^p$ matrix with precisely one $1$ and $d^p - 1 ~~0$'s 
in each row and column,
can have invariant subspaces.  In fact, it will be  block diagonal.  

\section {Properties of linearizing operators $X({\cal E},p)$}  \label{s:set}
 
\subsection{Kraus operator  form of $\Omega( {\cal E},p)$} \label{s:thetapf}

We first observe that conjugation of a tensor product
of operators by a shift operation induces a shift on the tensor product, e.g.,
  \begin{eqnarray}   \label{oplshft}
    L_p \Big[ B_1 \otimes B_2 \otimes \cdots \otimes B_p \Big] L_p^{-1} =
 B_2  \otimes \cdots \otimes B_p \otimes B_1.
 \end{eqnarray}
 More generally,
   \begin{eqnarray}  \label{opperm}
    \Pi_p \Big[ B_1 \otimes B_2 \otimes \cdots \otimes B_p \Big] \Pi_p^{-1} =
 B_{\Pi(1)}  \otimes B_{\Pi(2)}  \cdots  \otimes B_{\Pi(p)} .
 \end{eqnarray}
To prove (\ref{thetafin}), one can use (\ref{mappa}), and (\ref{oplshft}) 
to see that
 \begin{eqnarray}
 \big[\wh{\cal E}^{\otimes p}
(L_p ) \big] R_p & = & \bigg[ \sum_{k_1,\cdots,k_p} 
(A^{\dag}_{k_1} \otimes A^{\dag}_{k_2}
\otimes \cdots \otimes A^{\dag}_{k_p}) L_p 
(A_{k_1} \otimes A_{k_2}
\otimes \cdots \otimes A_{k_p}) \bigg] ~ L_p ^{-1} \nonumber  \\
 & = &   \sum_{k_1,\cdots,k_p} 
(A^{\dag}_{k_1}  \ot A^{\dag}_{k_2}
 \ot \cdots  \ot A^{\dag}_{k_p})   (A_{k_2} \ot A_{k_3}
 \ot \cdots   \ot A_{k_{p}} \ot A_{k_1})   \label{thetafinL} \\
 & = & \sum_{k_1,\cdots,k_p} 
A^{\dag}_{k_1} A_{k_2} \otimes A^{\dag}_{k_2} A_{k_3}
\otimes \cdots \otimes A^{\dag}_{k_p} A_{k_1}     \nonumber
\end{eqnarray}
which gives the desired result.  
Moreover, using a similar argument and (\ref{opperm}), one finds
\begin{eqnarray}
R_p  \big[\wh{\cal E}^{\otimes p} (L_p ) \big] =
& = & R_p \bigg[ \sum_{k_1,\cdots,k_p} 
(A^{\dag}_{k_1} \otimes A^{\dag}_{k_2}
\otimes \cdots \otimes A^{\dag}_{k_p}) L_p ^{-1}
(A_{k_1} \otimes A_{k_2}
\otimes \cdots \otimes A_{k_p}) \bigg]  \nonumber  \\
 & = &   \sum_{k_1,\cdots,k_p} 
A^{\dag}_{k_p} A_{k_1} \otimes A^{\dag}_{k_1} A_{k_2}
\otimes \cdots \otimes A^{\dag}_{k_{p-1}} A_{k_p}  .    \label{thetafinR}
 \end{eqnarray}
 Then by observing that both (\ref{thetafinL}) and (\ref{thetafinR}) involve
 tensor products of operators of the form  $A^{\dag}_{k_j} A_{k_{j+1}}$,
 one sees that after a change of variable in the summation indices,
e.g, $k_j \rightarrow  k_{j - 1}$ in  (\ref{thetafinR}), the two expression
are identical.  Therefore,  $R_P$ commutes with  $ \wh{\cal E}^{\otimes p} 
(L_p )$ and 
$ \Theta({\cal E},p) = \Omega({\cal E},p) \, R_p = R_p \Omega({\cal E},p) $.

\subsection{General permutations} \label{s:Xperm}

Define ${\cal X}({\cal E},p)$
the set of operators $X({\cal E},p)$ of ${\cal H}^{\otimes p}$
that satisfy the property~(\ref{prop1}) for all the input states $\gamma$
of~$\cal H$.  We have already seen that ${\wh{\cal E}}^{\ot p}(L_p)$ is in
${\cal X}({\cal E},p)$ which implies that it is non-empty.
Moreover, the linearity of Eq.~(\ref{prop1}) with respect to $X({\cal E},p)$
implies that  whenever $X ({\cal E},p)$ and $Y({\cal E},p)$ are in
${\cal X}({\cal E},p)$, then
$ a  X ({\cal E},p)+(1-a) Y({\cal E},p)$ is in $X({\cal E},p)$ is also.
This true for  any real number $a$ including $a < 0$ and $a > 1$,
and even for complex $a$.
By choosing $0 < a < 1$, we can also conclude that ${\cal X}({\cal E},p)$ 
is convex; however, ${\cal X}({\cal E},p)$  is not compact.
Because $\mbox{Tr} \big[{\cal E}(\gamma) \big]^p$ is real,  
\begin{eqnarray}
 \mbox{Tr} \, \gamma^{\ot p}  X({\cal E},p) = 
 \ovb{\mbox{Tr} \, \gamma^{\ot p} X({\cal E},p)} =
  \mbox{Tr} \big[ \gamma^{\ot p}   X({\cal E},p)\big]^{\dagger} =
\mbox{Tr}  \big[ X({\cal E},p)\big]^{\dagger}   \gamma^{\ot p} =
\mbox{Tr} \, \gamma^{\ot p}   \big[ X({\cal E},p)\big]^{\dagger}  
\end{eqnarray}
 for all density matrices $\gamma$.
    Therefore, whenever $X({\cal E},p)$ is in ${\cal X}({\cal E},p)$ 
so are $\big[ X({\cal E},p) \big]^{\dagger}$ and the self-adjoint operator
$\tfrac{1}{2} \Big( X({\cal E},p) + \big[ X({\cal E},p) \big]^{\dagger} \Big)$.

In view of the discussion in Appendix~\ref{s:trace} we can also
conclude that the operator
${\wh{\cal E}}^{\otimes p}(\Pi_p)$ is in $\cal X$ whenever
$\Pi_p$ is a permutation whose shortest cycle is length $p$.
Moreover, a modification of the argument in the preceding section shows that, 
for these permutations,
\begin{eqnarray} \label{Piomeg}
 {\wh{\cal E}}^{\otimes p}(\Pi_p)  \, \Pi_p^{\dag} =  \Pi_p^{\dag}  \, 
{\wh{\cal E}}^{\otimes p}(\Pi_p)  
=   \sum_{k_1,\cdots,k_p} 
A^{\dag}_{k_1} A_{\Pi(k_1)} \otimes A^{\dag}_{k_2} A_{\Pi(k_2)}
\otimes \cdots \otimes A^{\dag}_{k_p} A_{\Pi(k_p)}  \;.
\end{eqnarray}
Since $\Pi_p  \gamma^{\otimes p} \Pi_p^{\dagger} = \Pi_p$ for any permutation,
\begin{eqnarray}
\mbox{Tr}[\,\gamma^{\otimes p} \;( \Pi_p\, X({\cal E},p)\, \Pi_p^{\dag})\,] =
\mbox{Tr}[\, ( \Pi_p^{\dag} \gamma^{\otimes p} \Pi_p) \; X({\cal E},p) \,]  = 
  \mbox{Tr} \, \gamma^{\otimes p} X({\cal E},p)  
= \mbox{Tr}[{\cal E}(\gamma)^p] .
\label{setto}
\end{eqnarray}
Note that the map $P_p \mapsto \Pi_p P_p  \Pi_p^{\dag}$ does not
change the cycle structure of $P_p$, e.g, if $P_p$ is a product of a 
3-cycle and a disjoint
2-cycle, then so is $\Pi_p P_p  \Pi_p^{\dag}$.    
Thus, $\Pi_p L_p  \Pi_p^{\dag}$ 
is a permutation
whose shortest cycle is length $p$ 
irrespective of the cycle structure of $\Pi_p$.
One can show that
$\Pi_p \,\big[ {\wh{\cal E}}^{\otimes p}(L_p) \big] \, \Pi_p^{\dag} =  
     {\wh{\cal E}}^{\otimes p}(\Pi_p L_p  \Pi_p^{\dag})$,
with a similar result when $L_p$ is replaced by any permutation whose shortest
cycle is length $p$.

\subsection{Linearizing operators for pure inputs}  \label{s:Xpure}
The set ${\cal X}({\cal E},p)$ is a  subset of
 ${\cal X}_{\rm pure}({\cal E},p)$,  the set of operators,
which satisfy the property~(\ref{prop1}) when
$\gamma=|\psi\rangle\langle \psi|$ is pure.
We have already observed that   $\Theta({\cal E},p) = \Omega({\cal E},p) R_p$  
belongs to ${\cal X}_{\rm pure}({\cal E},p)$ but need not belong to 
${\cal X}({\cal E},p)$.  
It follows from (\ref{Piomeg}) that 
the operators ${\widehat{\cal E}}^{\otimes p}(\Pi_p)\Pi_p^{\dag}$ are
also in  ${\cal X}_{\rm pure}({\cal E},p)$.
In addition,  for any $X({\cal E},p) \in {\cal X}_{\rm pure}({\cal E},p)$ 
the operators
$X({\cal E},p) \, \Pi_p$ and $\Pi_p \, X ({\cal E},p)$ are also in 
${\cal X}_{\rm pure}({\cal E},p)$
for all permutations $\Pi_p$.
This follows from
\begin{eqnarray} 
   \mbox{Tr} \,\big[ \gamma \otimes 
\gamma \otimes \cdots \otimes \gamma \big] \, X({\cal E},p) \Pi_p 
& = & \mbox{Tr} \,  |\psi \rangle \langle \psi |\otimes |\psi \rangle 
\langle \psi | \otimes \cdots \otimes |\psi \rangle \langle \psi | \, 
X({\cal E},p) 
\Pi_p \nonumber \\
& = & \mbox{Tr} \, \Pi_p \, |\psi,\cdots,\psi \rangle
     \langle \psi,  \cdots ,\psi | \, X({\cal E},p) 
\nonumber \\
& = & \mbox{Tr} \, |\psi,  \cdots, \psi \rangle
     \langle \psi, \cdots, \psi | \, X({\cal E},p) 
\nonumber \\
& = & \mbox{Tr} \big[ \gamma \otimes \gamma \otimes \cdots \otimes 
\gamma \big] \, X({\cal E},p)= \mbox{Tr}{\cal E}(\gamma)^p \;, \nonumber
\end{eqnarray}
whenever $\gamma=|\psi\rangle\langle \psi|$ is pure.

\section{Operators for Werner-Holevo channel} \label{s:WH}

\subsection{General form of $\Omega({\cal W},p$)}\label{s:WH1}
It follows from (\ref{werner}), (\ref{shftrep}) and (\ref{omega})
that for the WH channel,
\begin{widetext} \begin{eqnarray}
\Omega({\cal W},p) & = & \sum_{\xi_1 \cdots \xi_p}
   {\cal W}\big( |\xi_2 \rangle \langle \xi_1| \big) \otimes
     {\cal W}\big( |\xi_3 \rangle \langle \xi_2| \big) \otimes
     \cdots {\cal W}\big( | \xi_p \rangle \langle \xi_{p-1}| \big)
     {\cal W}\big( |\xi_1 \rangle \langle \xi_p| \big) \nonumber \\
     & = & \frac{1}{(d-1)^p}\sum_{\xi_1 \cdots \xi_p}
       \big( \delta_{\xi_2 \xi_1} \openone - |\ovb{\xi}_1 \rangle \langle
      \ovb{\xi}_2| \big)
      \otimes \big( \delta_{\xi_3 \xi_2} \openone- |\ovb{\xi}_2
      \rangle \langle \ovb{\xi}_3| \big)
      \otimes \cdots \otimes
      \big( \delta_{\xi_1 \xi_p}\openone -
      |\ovb{\xi}_p \rangle \langle \ovb{\xi}_1| \big)
      \nonumber \\
      & = & \frac{1}{(d-1)^p} \bigg[ d\openone - \bigg( \sum_{\xi_1}
      |\ovb{\xi}_1 \rangle \langle \ovb{\xi}_1| +
          \sum_{\xi_2} |\ovb{\xi}_1 \rangle \langle \ovb{\xi}_2| + \cdots +
           \sum_{\xi_p} |\ovb{\xi}_1 \rangle \langle \ovb{\xi}_p| \bigg)
       \nonumber \\
       & ~ & +
       \sum_{a < b} \bigg( \sum_{\xi_a \xi_b} |\ovb{\xi}_a \ovb{\xi}_b \rangle
       \langle \ovb{\xi}_b \ovb{\xi}_a | \bigg) - \cdots
       + (-1)^p \sum_{\xi_1 \cdots \xi_p}
       | \ovb{\xi}_1 \ovb{\xi}_2 \cdots
       \ovb{\xi}_{p-1} \ovb{\xi}_p \rangle
       \langle \ovb{\xi}_2 \ovb{\xi}_3 \cdots \ovb{\xi}_p
       \ovb{\xi}_1 | \label{WHmid} \\
       & = &
       \frac{1}{(d-1)^p} \bigg[ (d-p) \openone +
       \sum_{a < b} S_{ab} - \sum_{a < b < c} R_3(a,b,c)
       + \cdots + (-1)^p R_p \bigg] \label{WHgen}
\end{eqnarray}
\end{widetext}
where we have used the notation introduced at the end of
Appendix~\ref{s:shift}.
Note that the orthonormal basis $\{ |\xi_j \rangle \}$ can be chosen real,
but even if it is not, $\{ | \ovb{\xi}_j \rangle \}$
gives another orthonormal basis for ${\cal H}$
for which the representation (\ref{shftrep}) is also valid.

It is useful to compare the structure of (\ref{WHgen}) to that of a
binomial expansion. The term in square brackets is a sum of shift
operators $R_k$ of order $k=0,1,2,\cdots p$. For $k\geqslant 2$ the
number of $R_k$ is $\binom{p}{k}$ with coefficient $(-1)^k$.
In view of (\ref{WHmid}), the $(d-p) \openone$ term should be regarded
as the sum of a $k=0$ term $d\openone$ and a $k=1$ term $-p \openone$.
The coefficient of the $k=0$ term is anomalous, since it has the value $d$
rater than $1$. This implies that the row and column sums of the matrix
representing $\Omega({\cal W},p)$ in the orthonormal basis
$\{|\ovb{\xi}_{j_1}\ovb{\xi}_{j_2} \cdots \ovb{\xi}_{j_p}\rangle\}$ of
${\cal H}^{\ot p}$ are
\begin{eqnarray} \label{WHrowsum}
\frac{1}{(d-1)^p} \bigg[ d + \sum_{k = 1}^{p} (-1)^k \binom{p}{k} \bigg]
%=\frac{1}{(d-1)^p} \bigg[ d-1 + \sum_{k = 0}^{p} (-1)^k \binom{p}{k} \bigg]
= \frac{d-1}{(d-1)^p} \;.
\end{eqnarray}
We similarly find that the sum of the absolute values of elements in any
row or column sum is bounded above by
\begin{eqnarray}
\frac{1}{(d-1)^p} \bigg[ d + \sum_{k = 1}^{p}
\binom{p}{k} \bigg] = \frac{\big(d-1 + 2^p \big)}{(d-1)^p}
\nonumber \;,
\end{eqnarray}
and we will use the fact 
that $\displaystyle{ \sum_{k = 2}^{p} \binom{p}{k} = 2^p - p - 1}$.

\subsection{Singular value analysis for $p = 3$}\label{s:p3}

We first remark that one can reduce the analysis of 
$\Omega({\cal W},3)$ to that of its 
$6 \times 6$ blocks without using   Lemma~\ref{bigblk}.
When $p = 3$,  all blocks with basis vectors $|jjk \ket$
with $j \neq k$ have only non-negative
elements.  To see why, note that  the only negative 
contribution comes from $R_3$, for which
$\langle jjk | R_3 | jkj \rangle = -1 $ is the only 
non-zero element of the row corresponding to $jjk$.  But  
$\langle jjk | \Omega  | jkj \rangle 
\geqslant  \langle jjk | (S_{ac} - R_3)  | jkj \rangle = 0$.
Therefore, every $3 \times 3$ blocks is represented by  a
stochastic matrix and, hence, its column sum $(d-1)^{1-p}$ is also
its largest singular value.
Thus, only the $6 \times 6$ blocks of $\Omega({\cal W},3)$  
 can have negative elements
and, hence, a singular value greater than $(d-1)^{1-p}$.

  Using an ordered basis whose first three elements are
 $\{ |ijk \rangle, L_3 |ijk \rangle, L_3^2 |ijk \rangle\} $ and last three
 $S_{ab}\{ |ijk \rangle,S_{bc} |ijk \rangle, S_{ac} |ijk \rangle \}$,
 one can write
 each $6 \times 6$ block as
 $ (d-1)^{-3} F$ with 
 \begin{eqnarray} F = (d-3) \openone_6 +  
 \begin{pmatrix} - L_3 & V \\ V& -L_3 \end{pmatrix}  \qquad
\hbox{and} \qquad   V = \begin{pmatrix} 1 & 1 & 1 \\ 1 & 1 & 1 \\ 1 & 1 & 1 \end{pmatrix}. 
\end{eqnarray}
Then
\begin{eqnarray}
 F^{\dag} F & = & (d-3)^2 \openone + (d-3) \big[G + G^{\dag} \big] +
G^{\dag}G \nonumber \\
        & = & (d^2 -5d +7) \openone_6 +
           \begin{pmatrix} -d+6 & 2d-8 \\ 2d-8 & -d+6 \end{pmatrix} \ot V .
\label{omeg3pfin}
\end{eqnarray}
    Since the eigenvalues of $V$ are $ 3,0,0$,
 the non-zero eigenvalues of $F^{\dag} F$ are $d^2 -5d +7 $
 (with 4-fold degeneracy) and
 $(d^2 -5d +7) + 3[(6-d) \pm (2d-8)]$ or
$(d-7)^2$ and $(d-1)^2$. Now
$(d^2 -5d +7) \leqslant (d-1)^2$ when $d \geqslant 2$ and
$(d-7)^2 \leqslant (d-1)^2$ if and only if
$d \geqslant 4$. Therefore, when $d \geqslant 4$
the largest singular value of this block is  $d-1$  which implies that
 the largest singular value of  $\| \Omega({\cal W},3) \|_{\infty} = (d - 1)^{-2}$.

\subsection{Singular value analysis for $p = 4$} \label{s:mathematica}

For $p = 4$, one  can show that the $4 \times 4$ and $6 \times 6$  blocks have
only non-negative elements. Therefore, their largest singular value is the
same as the column sum $(d-1)^{-3}$.
$\Omega({\cal W},4)$ also has $12 \times 12$ blocks corresponding to permutations of
$  |ijkk\rangle$,  with $i,j,k$ distinct 
and  $24 \times 24$ blocks  corresponding to permutations of
$ |ijk \ell \rangle$,  with $i,j,k, \ell$ distinct.
By Lemma~\ref{bigblk}, the largest singular value is  
associated with the latter.   Nevertheless, an analysis of all blocks 
was performed using Mathematica, yielding the  results  
summarized in Table~\ref{dec}.   This confirms that the largest
singular value of  $\Omega({\cal W},4)$ is $(d - 1)^{-3}$ when $ d \geq 8$.
\begin{table}[h!]
\begin{tabular}{|c||c|c|}
\hline \hline
singular value & degeneracy & degeneracy \\
$\times (d-1)^4$ & ($12 \times 12$ blocks) & ($24 \times 24$ blocks)  \\ 
\hline
$\sqrt{d^2-12 d + 45} $ &$2$ &$6$\\
$|d-5|$  &$1$ &$3$  \\
$|d-3|$  &$3$ &  $5$ \\
$\sqrt{d^2 -4d +5}$ & $4$ & $6$  \\
$|d-1|$ & $2$ & $3$ \\ 
$|d-15|$ &$0$ &$1$ \\
\hline \hline  
\end{tabular} 
\caption{Singular value decomposition of $\Omega({\cal W},4)$ on 
the twelve dimensional subspace generated by the vectors 
$\{ |ijkk\rangle, |jikk\rangle, 
\cdots,|kjik\rangle\}$ and the twenty four dimensional subspace
generated by  $\{|ijk \ell \rangle, |jik \ell\rangle, 
\cdots,|kji \ell\rangle\}$.
The singular  values of $\Omega({\cal W},4)$
are given in the left column, with the 
corresponding degeneracies in the central and right columns.
\label{dec}} \end{table}

\subsection{Structure of largest block} \label{app:irrep}

\subsubsection{Preliminaries}

Recall that every permutation $P$ in ${\cal S}_p $ can be 
classified as even or odd, depending on the number of
transpositions  (or SWAP) operators needed
to write it as a product 
$P = S_{a_1b_1} S_{a_2b_2} \ldots S_{a_mb_m}$.   Although
this decomposition is not unique, $m$ is either always  even
or  always odd.    Let $|P|$ be the minimal number of swaps
needed so that $(-)^{|P|} =  \begin{cases} +1 & \text{if $P$ is even} \\
  -1 & \text{if $P$ is odd} \end{cases}$.
Note that  $S(a,b)$ and $R_4(a,b,c,d)$ are odd and
$R_3(a,b,c)$ is even.   More generally, a shift of $j$ elements
is even when $j$ is odd and odd when $j$ is even.   Thus,
one can write 
\begin{eqnarray}  \label{omegeo}
  \Omega({\cal W},p) =  \frac{1}{(d-1)^p} \Big[ (d-p) \openone +  \wt{\Omega}_{\rm odd} 
     -  \wt{\Omega}_{\rm even}  \Big]
\end{eqnarray}
where $ \wt{\Omega}_{\rm odd} $ is the sum over odd permutations
(even shifts) in \eqref{WHgen} and 
$  \wt{\Omega}_{\rm even} $  the
sum over even permutations (odd shifts) in \eqref{WHgen}.

Fix $k_1 < k_2 < \ldots < k_p$  and let  ${\cal K}$ denote the 
subspace spanned by  
$\{ P  | \xi_{k_1}, \xi_{k_2} , \ldots \xi_{k_p} \ket : P   \in {\cal S}_p  \} $
where  $| \xi_k \ket$ is an orthonormal basis for 
${\mathbb C}^d$ and
the action of $P$ is as defined in (\ref{permH}).
The matrix  representing a particular permutation
 operator $\Pi$ has elements
 \begin{eqnarray}   \label{pimat}
    \pi_{st} = \bra    \xi_{k_1}, \xi_{k_2} , \ldots \xi_{k_p}| \, P_s^{\dag} 
      \Pi    P _t  \, | \xi_{k_1}, \xi_{k_2} , \ldots \xi_{k_p} \ket 
 \end{eqnarray}
which depends {\em only} on the labeling $P_s, ~s = 1,2 \ldots p!$
of elements of ${\cal S}_p$ and not on the choice of indices
$k_j$ or vectors $\xi_j$.    It will be convenient to simply
use $| k \ket$ to denote $| \xi_k \ket$, and to write
$| \Pi( k_1, k_2, \ldots k_p) \ket$ for    
$ \Pi | \xi_{k_1}, \xi_{k_2} , \ldots \xi_{k_p} \ket $.  (The condition $k_j < k_{j+1}$ is 
only a convenient convention; the essential requirement is
that  the $k_j$ are distinct.)   

\subsubsection{Irreducible representation structure}

The matrix representing the action of a permutation $\Pi$ on the
vectors $ \{ P   | {k_1}, {k_2} , \ldots {k_p} \ket : P \in {\cal S}_p \}$ is  identical to  
its matrix in the regular representation of ${\cal S}_p$.   Therefore, one can
find a unitary  transformation to a basis whose components
form disjoint subsets which transform as the irreducible representations of
${\cal S}_p $.    This basis change simultaneously converts all permutations
to a block diagonal form.   Thus,  $\Omega({\cal W},p) $, is also
block diagonal with each block 
corresponding to an irreducible representation of ${\cal S}_p $.
The two one-dimensional representations, therefore, yield eigenvectors
of $  \Omega({\cal W},p)   $.  In fact
 \begin{subequations}  \label{evalirred} \begin{eqnarray} 
     \Omega({\cal W},p) |\phi_{\rm sym} \ket  & = &  ~~  \frac{d-1}{(d-1)^p} ~|\phi_{\rm sym} \ket  \\
      \Omega({\cal W},p) |\phi_{\rm anti} \ket  & = &  \frac{d - 2^p + 1}{(d-1)^p} \, |\phi_{\rm anti} \ket
\end{eqnarray}  \end{subequations}  where
 \begin{subequations}  \begin{eqnarray}    
     |\phi_{\rm sym} \ket  ~ = &   \tfrac{1}{\sqrt{p!}} 
      \displaystyle{\sum_{P   \in {\cal S}_p}} |P ( {k_1}, {k_2} , \ldots {k_p}) \ket 
       & =   ~ \tfrac{1}{\sqrt{2}} \Big( |u_{\rm even} \ket +  |u_{\rm odd} \ket \Big) \\
      |\phi_{\rm anti} \ket ~ = &   \tfrac{1}{\sqrt{p!}}  \displaystyle{\sum_{P   \in {\cal S}_p}}
             (-)^{|P| } |P ( {k_1}, {k_2} , \ldots {k_p}) \ket
             & =  ~ \tfrac{1}{\sqrt{2}}\Big( |u_{\rm even} \ket -  |u_{\rm odd} \ket \Big) 
             \end{eqnarray}   \end{subequations} 
with  $|u_{\rm even} \ket = 
\sqrt{\tfrac{2}{p!}} \displaystyle{\sum_{ P_ {\rm even}} } |P   ( {k_1}, {k_2} , \ldots {k_p} )\ket $, and
$|u_{\rm odd} \ket = 
\sqrt{\tfrac{2}{p!}}\displaystyle{\sum_{ P_ {\rm odd}} }|P  ( {k_1}, {k_2} , \ldots {k_p}) \ket $.
If we could conclude that the largest singular value of $(d-1)^p  \Omega({\cal W},p)$ is 
associated with a one-dimensional representation of ${\cal S}_p $, then 
we could conclude that $(d-1)^p \|  \Omega({\cal W},p) \|_{\infty} = 
 \max\{d-1, |d - 2^p +1| \}$.     Note that this maximum is clearly $d - 1$ when $d \geq 2^p -1 $.
 For $d < 2^p $,  the maximum is  $d - 1$  if and only if
 $2^p - d - 1 \leq d - 1 ~ \Leftrightarrow ~2d \geq 2^p$.

\subsubsection{Odd/even structure}

We now describe the odd/even structure of $\Omega({\cal W},p)$.
We can divide the $p!$ basis vectors of ${\cal K}$ into two equal subsets, those
 of the form $ P_{\rm even} | k_1, k_2, \ldots k_p \rangle$ and  
those of the form $ P_{\rm odd} |k_1, k_2, \ldots k_p\rangle $.     We will
denote their spans as ${\cal K}_{\rm even}$  and ${\cal K}_{\rm odd}$ respectively.
Now $ \langle  k_1, k_2, \ldots k_p |  \Pi | k_1, k_2, \ldots k_p \rangle = 0$ unless 
$\Pi$ is the identity permutation.
Therefore 
$\langle  P_s (k_1, k_2, \ldots k_p) |  \Pi |  P_t (k_1, k_2, \ldots k_p) \rangle = 0$ 
unless $\Pi =  P_s  P_t^{\dag} = I$.
Moreover, since the identity is an even permutation
\begin{eqnarray}
    \langle  P_{\rm even} (k_1, k_2, \ldots k_p )  
|  \Pi_{\rm odd} | \wt{P}_{\rm even} (k_1, k_2, \ldots k_p) \rangle = 
       \langle  P_{\rm odd} (k_1, k_2, \ldots k_p)  
|  \Pi_{\rm odd} | \wt{P}_{\rm odd} (k_1, k_2, \ldots k_p) \rangle  = 0 \\
       \langle  P_{\rm even} (k_1, k_2, \ldots k_p)  
|  \Pi_{\rm even} | \wt{P}_{\rm odd} (k_1, k_2, \ldots k_p )\rangle = 
       \langle  P_{\rm odd} (k_1, k_2, \ldots k_p )  
|  \Pi_{\rm even} | \wt{P}_{\rm even} ( k_1, k_2, \ldots k_p) \rangle  = 0  
\end{eqnarray}
Thus, the  largest block of $(d-1)^p\Omega({\cal W},p)$
can be written in the  form
$B = (d-p) \openone +  
  \begin{pmatrix} - B_{\rm ee} & B_{\rm eo}  \\  B_{\rm oe} & -B_{\rm oo} \end{pmatrix} $
  with $B_{\rm ee}$ and $B_{\rm oo}$ determined by  $\wt{\Omega}_{\rm even} $
  and $B_{\rm eo}$ and $B_{\rm oe}$ determined by $ \wt{\Omega}_{\rm odd}$.

It is useful to relate the order of elements  within the bases associated with
 odd and even permutations.
Let $P_1, P_2, \cdots P_M$  with $M = {p!/2}$ denote the even permutations 
(with $P_1 = \openone$)
and $P_{t+ M} =   P_t S$
the odd, where $S$ denotes the swap operator 
$S(k_1, k_2, k_3 \ldots k_p)  =  k_2, k_1, k_3 \ldots k_p$.  
(There is nothing special about applying SWAP
to the first two elements.  Any fixed choice would do.)
Then
\begin{eqnarray}
     b_{s ,  t+M} & = &  \bra P_s(k_1, k_2, \ldots k_p ) | \,  \Pi  \, | P_t S(k_1, k_2, \ldots k_p ) \ket \nonumber \\
        & = &  \bra P_s S (k_2, k_1, \ldots k_p ) | \,  \Pi  \, | P_t ( k_2, k_1, \ldots k_p ) \ket \\
        & = &  b_{s+M ,  t} \nonumber.
\end{eqnarray}
 where we used the fact that the matrix representing a permutation is
 independent of the initial choice of $k_i$.    Thus, $B_{\rm eo} = B_{\rm oe}$
 and, for the same reason, $B_{\rm ee} = B_{\rm oo}$, and we can write 
  \begin{eqnarray}
B = (d-p) \openone + 
    \begin{pmatrix} - W_{\rm e} & W_{\rm o}  \\  W_{\rm o}   &   - W_{\rm e} \end{pmatrix} = 
   (d-p) \openone + B_{\rm off} .
\end{eqnarray}
where $W_{\rm e} $ and $W_{\rm o}$ are determined by $\wt{\Omega}_{\rm even} $
and $\wt{\Omega}_{\rm odd} $ respectively.
By conjugating with
$\begin{pmatrix} \openone   & 0 \\ 0 & - \openone  \end{pmatrix} $, one finds that
$B$  has the same singular values as  
  \begin{eqnarray}   \label{Fpequiv}
G =  (d-p) \openone      
       -   \begin{pmatrix}   W_{\rm e} & W_{\rm o}  \\  W_{\rm o}   &     W_{\rm e} \end{pmatrix}  =   
       (d-p) \openone -  G_{\rm off}
  \end{eqnarray}
%  All diagonal elements $a_{jj}$ are zero and all the
 %  diagonal elements of $B$ and $G$ are simply $d-p$.   
  Notee that we have shown that the restriction of $ \Omega({\cal W},p)$ 
  to ${\cal K}$ is similar to
$     \frac{1}{(d-1)^p} \Big[ (d-p) \openone -  \wt{\Omega}_{\rm odd} 
     -  \wt{\Omega}_{\rm even}  \Big] $
which differs from \eqref{omegeo} by a sign.   Although this may seem
surprising, it could easily be established directly by observing that
any vector $| v \ket \in {\cal K}$ can be written as
 $| v \ket = | v_{\rm even} \ket +  | v_{\rm odd} \ket $ with  
 $| v_{\rm even} \ket  \in {\cal K}_{\rm even}$  and 
$ | v_{\rm odd} \ket  \in {\cal K}_{\rm odd}$.      
Using \eqref{WHrowsum} and related combinatorics,  
 one finds that the row and column sums
  of $B, G, W_{\rm e}$ and $ W_{\rm o} $ are, respectively, 
  $d - 1 , \linebreak d- 2^p +1, ~ 2^{p-1} - p$, and $ 2^{p-1} - 1$. 
   It   follows  
that  $ d - 1$ and  $d - 2^p + 1$ are eigenvalues of $B$ and $G$.  
 \begin{subequations} \label{evalblk} \begin{eqnarray}   
   B  \begin{pmatrix} 1 \\ 1 \end{pmatrix}   = (d \! - \!1)   \begin{pmatrix} 1 \\ 1 \end{pmatrix},   & ~ &
   B  \begin{pmatrix} ~1 \\ -1 \end{pmatrix}   = (d \! - \! 2^p \! + \!  1)   \begin{pmatrix} ~1 \\ -1 \end{pmatrix}, \\
   G  \begin{pmatrix} ~1 \\ -1 \end{pmatrix} = (d\! - \!1)   \begin{pmatrix} ~1 \\ -1 \end {pmatrix},   & ~ & 
   G  \begin{pmatrix} 1 \\ 1 \end{pmatrix}   = (d \! - \! 2^p \! + \!   1)  \begin{pmatrix} 1 \\ 1 \end{pmatrix},
     \end{eqnarray}     \end{subequations} 
    where $1$ denotes a vector with all $1$'s.    These are easily seen to
    be equivalent to \eqref{evalirred}. 

The main reason for changing $B$ to the form \eqref{Fpequiv}
is that $ G_{\rm off}$ is a multiple of a double stochastic matrix, 
its column sum $2^p -  p-1$ is both  its largest
eigenvalue and its largest singular value.   Therefore, $d - 2^p + 1$
is the smallest eigenvalue of $G$; however,  even when  
it is the most negative eigenvalue, 
we cannot conclude that it is also  the largest
singular value because $G$ could have a positive, or complex,  eigenvalue of greater
magnitude.

\noindent {\bf Remark:} Conjugating $B$ with the block Hadamard transform 
$H =  \tfrac{1}{\sqrt{2}}\begin{pmatrix} \openone   &  \openone \\    \openone &  -\openone \end{pmatrix} $
corresponds to making the change of basis to (\ref{eq:block12}).   One finds
 \begin{eqnarray}   \label{had}
H F H^{\dag} = (d-p) \openone +  
 \begin{pmatrix} -W_{\rm e} +  W_{\rm o}  & 0  \\  
  0 & -W_{\rm e} -  W_{\rm o}  \end{pmatrix} .
 \end{eqnarray}
% Thus, $(d-1)^p \wt{B}_{++} = -W_{\rm e} +  W_{\rm o} $.  

\subsection{Proof that  $\| \Omega({\cal W},p)  \|_{\infty} $ is attained
on the largest blocks}  \label{app:bigblk}

As above,    
fix $k_1 < k_2 < \ldots < k_p$  and let $B$ denote the block of
$\Omega({\cal W},p)$ corresponding to their span ${\cal K}$.

For simplicity, we first compare
the singular values of $B$ to those for a block spanned by vectors of the form
\begin{eqnarray} \label{basjj}
\big\{  \Pi | j,j, {k_3}, \ldots , {k_p} \ket : \Pi  \in {\cal S}_p \big\} 
\end{eqnarray}
with $j < k_3 < \ldots < k_p$.     Observe that
\begin{eqnarray}  \label{eq:block12}
\Big\{ \tfrac{1}{\sqrt{2}} \Pi \big( | k_1, k_2, k_3 , \ldots k_p  \ket 
   \pm   |  k_2, k_1, k_3 , \ldots k_p  \ket  \big): 
     \Pi  \in {\cal S}_p, ~\Pi  \neq S_{12}  \Big\}  
\end{eqnarray}
is another orthonormal basis for ${\cal K}$, and   let $V$ be
the unitary matrix for the basis change from   
$\{ P| k_1,k_2, \ldots k_p  \ket , \,  P \in {\cal S}_p\} $ to
(\ref{eq:block12}).    Let ${\cal K}_{\pm}$ denote the subspace
spanned by vectors with a $\pm$ sign in (\ref{eq:block12}),
and   $\wt{B}_{++}$   the
submatrix for the restriction of $V B V^{\dag}$ to the subspace
${\cal K}_{+}$.       The effect of any
permutation on vectors of the form (\ref{basjj}) and
those with a $+$ sign in \eqref{eq:block12} is the same. 
Therefore, $\wt{B}_{++}$ is identical to the matrix for the  
restriction of $ \Omega({\cal W},p)$ to the span 
of (\ref{basjj}), and  
    the largest singular value of the latter is the same as
\begin{eqnarray}
  \| \wt{B}_{++} \|_{\infty}  & = & \sup_{\phi \in {\cal K}_+}
  \frac{ \bra \phi, \wt{B}_{++}^{\dag}  \wt{B}_{++}  \phi \ket}{ \| \phi \|^2}   
   = \sup_{\phi \in {\cal K}_+}   \nonumber
  \frac{ \bra \phi, V B^{\dag} B  V^{\dag} \phi \ket}{ \| \phi \|^2} \\  \label{bigsv}
   & \leq & \sup_{\phi \in {\cal K}} 
       \frac{ \bra \phi, V B^{\dag} B  V^{\dag} \phi \ket}{ \| \phi \|^2} = \|B \|_{\infty}^2 .
\end{eqnarray}
In  \eqref{had} we showed that
$\wt{B}_{++}  = -W_{\rm e} +  W_{\rm o} $ and that $B$ is block diagonal,
which immediately implies that the singular values of $\wt{B}_{++} $
are a subset of those for  $B$.   This is stronger than  \eqref{bigsv},
but does not  necessarily generalize.

Next, consider a block for a subspace spanned by vectors of the form
\begin{eqnarray} \label{basjm}
\big\{   \Pi | j, j, \ldots, j, k_{m+1}, \ldots k_p \ket : \Pi  \in {\cal S}_p \big\} 
\end{eqnarray}
with $m$ occurences of $j$ and $j < k_{m+1} \ldots k_p$.    
We adopt the
convention that  $Q  \in {\cal S}_m$ denotes a permutation of $\{ 1, 2, \ldots, m \}$.
Choose $p!/m!$ permutations $P_t \in {\cal S}_p $ such that each
$P_t$ is in a distinct coset of ${\cal S}_p/{\cal S}_m$ or, equivalently
$P_s P_t^{-1} \notin {\cal S}_m~ \forall \, s \neq t$.  Then the vectors
\begin{eqnarray}  \label{eq:msym}
  |\phi_t \ket =  \tfrac{1}{\sqrt{m!}}   \displaystyle{\sum_{Q  \in {\cal S}_m}}  
  P_t Q   | k_1 \ldots  k_m , k_{m+1}, \ldots k_p \ket .
\end{eqnarray}
transform under permutations exactly as those in \eqref{basjm}.
Therefore, the restriction of $B$ to the span of \eqref{eq:msym}
 is represented by the same matrix  as the block of 
$\Omega({\cal W},p)$ corresponding to (\ref{basjm}).   Then, as in
(\ref{bigsv}), its largest singular value is bounded above by  $\|B \|_{\infty}$.

To deal with the general case, note that the restriction 
$j < k_{m+1} < k_{m+2} < \ldots < k_p$ does not play an essential role.   
The same argument works whenever $j$ is distinct from the remaining
$k_i$ with $i > m$.    Then, for example, the
\begin{eqnarray*}
\lefteqn{  \hbox{largest singular value of the block for permutations of } ~ |i,i,i,j,j,k_6 \ldots k_p \ket}
  \\ & \leq & 
   \hbox{largest singular value of the block for permutations of } ~ |i,i,i,k_4,k_5, \ldots k_p \ket
    \\ & \leq & 
   \hbox{largest singular value of the block for permutations of } ~ |k_1, k_2,  \ldots k_p \ket
~   =  ~ \|B \|_{\infty}.
\end{eqnarray*}
Proceeding in this way, one can complete the argument by induction.   
Alternatively, one could consider cosets for repeated indices, such as
${\cal S}_p/({\cal S}_3 \times {\cal S}_2)$ in this example.

\bigskip

\acknowledgments
 The contribution of V.G. to this paper was funded by
the European Community under contracts IST-SQUIBIT,
IST-SQUBIT2, and RTN-Nanoscale Dynamics. 
V.G. would like to thank P. Zanardi for comments and criticism.  V.G. and 
M.B.R. are grateful to
M. D'Ariano for the opportunity of participating in the Quantum Information 
Processing workshop in Pavia, Italy.   The contributions of M.B.R. were
 partially supported  by  the National Security Agency (NSA) and 
Advanced Research and Development Activity (ARDA) under 
Army Research Office (ARO) contract number  
DAAD19-02-1-0065, and by the National Science Foundation under Grant  DMS-0314228.


\begin{references}

\bibitem{AB}K.~M.~R. Audenaert and S.~L. Braunstein,
``On strong superadditivity of 
the entanglement of formation'', {\em Commun. Math. Phys.}  
{\bf 246}, 443--452 (2004).

\bibitem{AF}R. Alicki and M. Fannes, 
``Note on multiple additivity of minimal entropy output of extreme 
$SU(d)$-covariant channels''
eprint quant-ph/0407033.

\bibitem{AH}G.~G. Amosov and A.~S. Holevo,
``On the multiplicativity 
conjecture for quantum channels'', 
{\em Theor. Probab. Appl.} {\bf 47}, no.1, 143--146 (2002).

\bibitem{AHW}  G.~G. Amosov, A.~S. Holevo, and R.~F. Werner, 
``On Some Additivity Problems in Quantum Information Theory'', 
{\em Problems in Information Transmission}, 
{\bf 36}, 305 -- 313 (2000).  eprint  math-ph/0003002.


\bibitem{BFS} C. H. Bennett C. A. Fuchs, and J. A. Smolin, 
``Entanglement-enhanced classical communication on a noisy quantum channel''
in {\em Quantum Communication, Computing and Measurement} O. Hirota, A.S. Holevo,
and C. M. Caves, Eds. (New York, Plenum, 1997), pg. 79.

\bibitem{SHOR} C. H. Bennett and P. W. Shor, 
``Quantum Information Theory''
{\em IEEE Trans. Inf. Theory} {\bf 44}, 2724 (1998).

\bibitem{CAVES} C. Caves and K. W\'odkiewicz, 
``Classical Phase-Space Descriptions of Continuous-Variable Teleportation''
{\em Phys. Rev. Lett.} {\bf 93}, 040506 (2004).

\bibitem{Choi} M-D Choi,
``Completely Positive Linear Maps on Complex Matrices'' 
{\em Lin. Alg. Appl.} {\bf 10}, 285--290 (1975).

\bibitem{Davies} B. Davies,
{\em Quantum Theory of Open Systems} 
(Academic Press, 1976).

\bibitem{DHS}
N. Datta, A. S. Holevo, and Y. Suhov, 
``A quantum channel with additive minimum output entropy''
eprint quant-ph/0403072

\bibitem{FA1} A. Fujiwara and P. Algoet, 
``One-to-one parametrization of quantum channels''
{\it Phys. Rev. A}  {\bf 59}, 3290--3294 (1999).

\bibitem{FA2} A. Fujiwara and T. Hashizum\'{e},
``Additivity of the capacity of   depolarizing channels''
{\it Phys lett. A} , {\bf  299}, 469--475 (2002).

\bibitem{GIOVA1}V. Giovannetti and S. Lloyd, 
``Additivity properties of a Gaussian channel''
{\em Phys. Rev. A},  {\bf 69}, 062307 (2004).

\bibitem{GIOVA2}V. Giovannetti, S. Lloyd, L. Maccone, J. H. Shapiro,
  and B. J. Yen, 
``Minimum R\'{e}nyi and Wehrl entropies at the output of bosonic channels''
{\em Phys. Rev. A} {\bf 70}, 022328 (2004). 


\bibitem{HWINFO}A. S. Holevo, 
``Some estimates of the information transmitted by 
quantum communication channel''
Probl. Inf. Trans. {\bf 9}, 177 (1973).

\bibitem{HJ1} R.A. Horn and C.R. Johnson, {\em Matrix Analysis}
(Cambridge University Press, 1985)

\bibitem{SVD} See also Section 2.1.10 of \cite{NC} and 
 Appendix A of \cite{KR1}.

\bibitem{King1} C. King,
``Maximization of capacity and p-norms for some product channels'',
{\em Journal of
Mathematical Physics}, 
{\bf 43}, no. 3,   1247 -- 1260 (2002).
  
\bibitem{King2} C. King,
``Additivity for unital qubit channels'',
{\em Journal of
Mathematical Physics}, {\bf 43}, no. 10 4641 -- 4653  (2002).

\bibitem{King3} C. King, ``The capacity of the quantum depolarizing channel'',
{\em IEEE Transactions on Information
Theory},   {\bf 49}, no. 1 221 -- 229, (2003).

\bibitem{King4} C. King, ``Maximal p-norms of entanglement breaking channels'',
{\em Quantum Information and Computation}, {\bf 3}, no. 2, 186 -- 190  (2003).

\bibitem{King5} C. King, ``An application of the Lieb-Thirring inequality 
in quantum information theory'', to appear in Proceedings of ICMP 2003,
quant-ph/0412046.
  
\bibitem{KNR}  C. King, M. Nathanson and M.~B. Ruskai,
  ``Multiplicativity results for entrywise positive maps'' {\em Lin. Alg. Appl.}
(in press) (2005), quant-ph/0409181.

\bibitem{KR1} C. King and M.~B. Ruskai, 
``Minimal Entropy of States Emerging from Noisy Quantum Channels'', 
{\em IEEE Trans. Info. Theory}, {\bf 47}, 192--209 (2001).
  
\bibitem{KR2} C. King and M. B. Ruskai, 
  ``Comments on multiplicativity of maximal p-norms when p = 2''
  in {\em Quantum Information, Statistics and Probability} 
  ed. by O. Hirota, pp. 102-114  (World Scientific, 2004)  quant-ph/0401026.
  
\bibitem{KRAUS} K. Kraus, Ann. Phys. {\bf 64}, 311 (1971); 
K. Kraus, {\em States,  Effects and Operations: Fundamental Notions of Quantum Theory} 
(Springer, Berlin, 1983).

\bibitem{MSW} K. Matsumoto, T. Shimono and A. Winter,
``Remarks on additivity of the Holevo channel capacity and of
 the entanglement of formation''
{\em Commun. Math. Phys.}  {\bf 246}, 427--442 (2004).

\bibitem{MY} K. Matsumoto and F. Yura, 
``Entanglement cost of antisymmetric states and additivity of capacity 
of some quantum channels'' {\em J. Phys. A},  {\bf 37}, L167 (2004).

\bibitem{NC} M. A. Nielsen and I. L. Chang, {\em Quantum
Computation and Quantum Information} (Cambridge University Press,
Cambridge, 2000).

\bibitem{Renyi}  A. R\'{e}nyi,  
``On measures of entropy and information'' pp. 547--561 in  
{\em Proc. 4th Berkeley Sympos. Math. Statist. and Prob.}, Vol. I 
(Univ. California Press, Berkeley, 1961).

\bibitem{RSW} M. B. Ruskai, S. Szarek, E. Werner, 
``An analysis of completely positive trace-preserving maps $M_2$''
{\em Lin. Alg. Appl.}  {\bf 347}, 159 (2002).

\bibitem{SEW} A. Serafini, J. Eisert, and M. M. Wolf, 
``Multiplicativity of maximal output purities of Gaussian channels 
under Gaussian inputs'' Phys. Rev. A {\bf 71}, 012320 (2005).

\bibitem{SHOREQ} P. W. Shor, 
 ``Equivalence of Additivity Questions in Quantum Information Theory'', 
{\em Commun. Math. Phys.}  {\bf 246}, 453-- 472 (2004). 

\bibitem{HW} R.~F. Werner and A.~S. Holevo, 
``Counterexample to an additivity conjecture 
for output purity of quantum channels'',
{\em  Jour. Math. Phys.} {\bf 43}, no. 9, 4353 -- 4357  (2002).

\bibitem{ZANNA} P. Zanardi and D. A. Lidar, 
``Purity and State Fidelity of Quantum Channels via Hamiltonians''
{\em  Phys. Rev. A}  {\bf 70}, 012315 (2004).


\end{references}
\end{document}